\PassOptionsToPackage{table}{xcolor}
\documentclass[manuscript,screen]{acmart}
\AtBeginDocument{%
  }

\usepackage{caption}
\usepackage{subcaption}
\usepackage{adjustbox}
\usepackage{mathtools}
\usepackage{multirow}
\usepackage{algorithm}
\usepackage{algpseudocode}
\usepackage{commath}
\usepackage{threeparttable}
\usepackage{graphicx}
\usepackage{tikz}
\usepackage{forest}
\usetikzlibrary{mindmap, shadows}
\usetikzlibrary{arrows,shapes,positioning,shadows,trees,arrows.meta}

\tikzset{
  basic/.style  = {draw, rounded corners=2pt, font=\small, thin, align=center, text width=8cm, drop shadow, rectangle}
}

\begin{document}

\title{Diffusion Models in Recommendation Systems: A Survey}

\author{Ting-Ruen Wei}
\email{twei2@scu.edu}
\affiliation{%
  \institution{Santa Clara University}
  \city{Santa Clara}
  \country{USA}
}

\author{Yi Fang}
\email{yfang@scu.edu}
\affiliation{%
  \institution{Santa Clara University}
  \city{Santa Clara}
  \country{USA}
}

\renewcommand{\shortauthors}{Wei et al.}

\begin{abstract}
Recommender systems remain an essential topic due to its wide application and business potential. Given the great generation capability exhibited by diffusion models in computer vision recently, many recommender systems have adopted diffusion models and found improvements in performance for various tasks. Research in this domain has been growing rapidly and calling for a systematic survey. In this survey paper, we propose and present a taxonomy based on three orthogonal axes to categorize recommender systems that utilize diffusion models. Distinct from a prior survey paper that categorizes based on the role of the diffusion model, we categorize based on the recommendation task at hand. The decision originates from the rationale that after all, the adoption of diffusion models is to enhance the recommendation performance, not vice versa: adapting the recommendation task to enable diffusion models. Nonetheless, we offer a unique perspective for diffusion models in recommender systems complementary to existing surveys. We present the foundational algorithms in diffusion models and their applications in recommender systems to summarize the rapid development in this field. Finally, we discuss open research directions to prepare and encourage further efforts to advance the field. We compile the relevant papers in a public GitHub repository\footnote{https://github.com/tingruew/DiffusionModels-In-RecSys}.
\end{abstract}





\maketitle

\section{Introduction}
Recommender systems recommend desired items to users and have gained popularity with the emergence of the World Wide Web. The internet gathers an abundance of users and has facilitated consumer domains such as e-commerce. Optimizing recommender systems leads to large profits for companies and better quality of life for their users. As one of the most common methods for optimizing recommender systems, collaborative filtering considers the similarity between users and items to deliver a recommendation for the target user. However, given the large set of users and items, a user often interacts with only a small portion of items, leading to sparsity \cite{chen2024data} in the dataset. Sparse datasets often pose a significant challenge for recommender systems, hindering their ability to deliver confident recommendations. To address the issue, some systems have incorporated generative adversarial networks \cite{goodfellow2014generative} to augment the dataset and increase its density \cite{gao2021recommender}. Nonetheless, generative adversarial networks are commonly known to suffer from training instability and mode collapse. In addition, the large and complex set of users and items in recommender systems introduce additional challenges in capturing intricate patterns. Given these challenges, there remains an interest in exploring alternative approaches, such as the diffusion model \cite{song2019generative, ho2020denoising}, that are better at modeling complex patterns without the shortcomings.

Recently, diffusion models in computer vision have exhibited these advantages. They possess the ability to model complex distributions and estimate noise. The concept first originated from non-equilibrium thermodynamics \cite{sohl2015deep} that aimed to model complex data distributions with probability distributions that are computationally tractable to learn and sample. The model gradually destroys a data distribution through an iterative forward diffusion process and produces a generative model that learns to restore the distribution through the reverse process. Later, NCSN \cite{song2019generative} proposed to perturb the data with Gaussian noise specifically and estimate the gradient of the perturbed distribution under all levels of noise. By sampling through Langevin dynamics, the generated images are comparable to those produced by generative adversarial networks. With the two previous works as foundation, Ho et al. \cite{ho2020denoising} developed Denoising Diffusion Probabilistic Models (DDPM) as another stream of generative models that rely on a fixed noise schedule. Together, they empowered recent breakthrough models in image generation, including Stable Diffusion \cite{rombach2022high}, DaLLE \cite{ramesh2021zero}, and DreamBooth \cite{ruiz2023dreambooth}. 

\begin{figure}[t]
    \centering
    \begin{minipage}{0.23\textwidth}
        \centering
        \includegraphics[width=\linewidth]{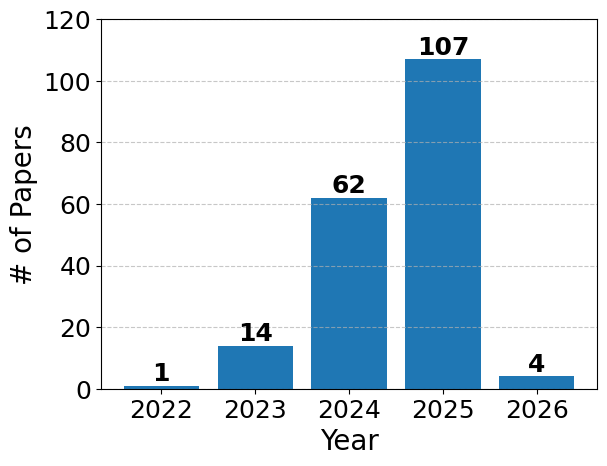}
        \caption{Growing Trend of Diffusion Models in Recommender Systems by November 2025 (non-cumulative).}
        \label{fig:trend}
    \end{minipage}\hfill
    \begin{minipage}{0.75\textwidth}
        \centering
        \begin{tikzpicture}[scale=0.8, >=Stealth, node distance=2cm]
    \draw[thick] (0, 0) -- (13.2, 0); 
    \foreach \x in {1, 3, 5, 7, 9, 10.7, 12.6} 
        \draw[thick] (\x, 0) -- (\x, 0.5);
        
    \foreach \x in {4, 6, 8, 10, 12} 
        \draw[thick] (\x, 0) -- (\x, -0.5);
    \node at (1, 0.7) [above, color=red!90, font=\scriptsize] {CODIGEM \cite{walker2022recommendation}};
    \node at (3, 0.7) [above, color=blue!90, font=\scriptsize] {DreamRec \cite{yang2024generate}};
    \node at (3, 1) [above, color=black!20!green, font=\scriptsize] {LD4MRec \cite{yu2023ld4mrec}};
    \node at (4, -0.7) [below, color=blue!90, font=\scriptsize] {DiffuRec \cite{li2023diffurec}};
    \node at (5, 0.7) [above, color=blue!90, font=\scriptsize] {Diff-POI \cite{qin2023diffusion}};
    \node at (6, -0.7) [below, color=cyan!90, font=\scriptsize] {DIEXRS \cite{guo2023explainable}};
    \node at (7, 0.7) [above, color=black!20!green, font=\scriptsize] {Diff-MSR \cite{wang2024diff}};
    
    \node at (8, -0.7) [below, color=cyan!90, font=\scriptsize] {IPDGI \cite{chen2023adversarial}};
    \node at (9, 0.7) [above, color=red!90, font=\scriptsize] {DGRM \cite{jiangzhou2024dgrm}};
    \node at (10, -0.7) [below, color=red!90, font=\scriptsize] {GiffCF \cite{zhu2024graph}};
    \node at (10, -1) [below, color=black!20!green, font=\scriptsize] {DiFashion \cite{xu2024diffusion}};
    \node at (10.7, 0.7) [above, color=black!20!green, font=\scriptsize] {DMSR \cite{tomasi2024diffusion}};
    \node at (10.7, 1) [above, color=cyan!90, font=\scriptsize] {CausalDiffRec \cite{zhao2024graph}};
    \node at (12, -0.7) [below, color=cyan!90, font=\scriptsize] {DifFaiRec \cite{jiang2024diffairec}};
    \node at (12.6, 0.7) [above, color=red!90, font=\scriptsize] {DiffuSAR \cite{zang2025diffusion}};
    \node at (12.6, 1) [above, color=blue!90, font=\scriptsize] {DCRec \cite{huang2024dual}};

    \node[below] at (0, 0) {\textbf{2022}};
    \node[below] at (2, 0) {\textbf{2023}};
    \node[below] at (6.5, 0) {\textbf{2024}};
    \node[below] at (13, 0) {\textbf{2025}};

    \draw[fill=red!90] (-0.5, -1.7) rectangle (0, -2.2); 
    \node at (0.1, -1.8) [right, font=\footnotesize] {Collaborative};
    \node at (0.1, -2.2) [right, font=\footnotesize] {Filtering};
    
    \draw[fill=blue!90] (2.5, -1.7) rectangle (3, -2.2); 
    \node at (3.1, -1.8) [right, font=\footnotesize] {Sequential};
    \node at (3.1, -2.2) [right, font=\footnotesize] {Recommendation};
    
    \draw[fill=black!20!green] (6, -1.7) rectangle (6.5, -2.2); 
    \node at (6.6, -1.8) [right, font=\footnotesize] {Data Modality};
    \node at (6.6, -2.2) [right, font=\footnotesize] {and Domain};
    
    \draw[fill=cyan!90] (9.5, -1.7) rectangle (10, -2.2); 
    \node at (10.1, -1.8) [right, font=\footnotesize] {Trustworthy Objectives};
    \node at (10.1, -2.2) [right, font=\footnotesize] {and Constraints};
    
    \end{tikzpicture}
    \caption{Timeline of Earliest Publications by Taxonomy Subcategory.}
    \Description{Timeline showing the emergence of various work in this domain.}
        \label{fig:timeline}
    \end{minipage}
\end{figure}

In general, diffusion models apply noise to the data in the forward process and train a neural network model to reconstruct the original data. At inference, the trained model predicts and removes an amount of noise from random noise in an iterative process to generate the final denoised output. While these models have exhibited appealing capabilities in image generation, the effectiveness stems from their ability to model complex distributions of the pixels. Given the data distribution, diffusion models have demonstrated their capabilities in capturing the underlying distribution and generating accordingly. Another advantage of diffusion models lies in their flexibility of the denoising network. While the U-net models \cite{ronneberger2015u} are common for image generation tasks, they can be replaced by other models, such as a multi-layer perceptron (MLP) or a transformer model \cite{vaswani2017attention}. The goal of the network is to serve as an approximator to identify the amount of added noise in the diffusion process. In addition, since diffusion models focus on optimizing the estimation of added noise, they do not experience the training instability that generative adversarial networks commonly have. Furthermore, the iterative denoising process, which incorporates random sampling from the standard normal distribution at each step, introduces variability and diversity in the final generated sample. This inherent randomness helps diffusion models avoid the issue of mode collapse, another common challenge of generative adversarial networks. Given the challenges in recommender systems and the advantages of diffusion models, they show great promise to enhancing recommender systems. Through the number of publications over recent years, we observe an upward trend of using diffusion models in recommender systems, as illustrated in Fig. \ref{fig:trend}.

Though NCSN \cite{song2019generative} and DDPM \cite{ho2020denoising} were introduced in 2019 and 2020, no work has adopted it for recommender systems until 2022. After that, there is an increasing amount of publication, representing a growing interest of this field, which motivates our efforts to construct this survey. We aim to provide a comprehensive scope of methods using diffusion models in recommender systems and describe the landscape, identifying gaps to be filled and encouraging research efforts. With the field developing at a fast pace, we aim to present the comprehensive knowledge for researchers who intend to quickly grasp the landscape and for graduate students who wish to enter this field.

\textbf{Survey Methodology.} With a focus on recommender systems involving diffusion models, we collect relevant papers using Google Scholar under queries such as ``diffusion models in recommender systems'' and ``diffusion models in recommendation.'' To differentiate from graph diffusion and information diffusion, we manually examine the definition of diffusion in each paper to only include papers using the diffusion model in common with NCSN \cite{song2019generative} or DDPM \cite{ho2020denoising}. Besides digging into Google Scholar until no relevant papers are available, we complement the paper collection strategy by manually checking the related work section of each relevant paper, ensuring complete coverage of relevant works. This yields 188 papers involving diffusion models in recommender systems that we include in this survey paper. Furthermore, Fig. \ref{fig:timeline} displays the timeline of some relevant papers categorized by our taxonomy.

\textbf{Related Work.} 
In recommender systems, surveys exist for fields such as reinforcement learning \cite{afsar2022reinforcement}, conversational recommender systems \cite{jannach2021survey}, graph neural networks \cite{wu2022graph, gao2023survey}, self-supervised learning \cite{yu2023self, ren2025comprehensive}, multimedia content \cite{deldjoo2020recommender, liu2024multimodal}, and session-based recommender systems \cite{wang2021survey}. These surveys focus on their particular subset of recommender systems and do not cover diffusion models. Several surveys have explored generative models in recommender systems \cite{nahta2024deep}, covering approaches such as variational autoencoders \cite{liang2024survey}, generative adversarial networks \cite{deldjoo2021survey}, and large language models (LLMs) \cite{wu2024survey, li2024survey, wang2024towards}. Additional studies have examined the role of LLMs in specific challenges such as cold-start scenarios \cite{zhang2025cold}, multimodal recommendation \cite{lopez2025survey}, while others have focused on the industrial applications of generative models in recommender systems \cite{xu2024survey}. In particular, Deldjoo et al. \cite{deldjoo2024review} mention diffusion models among generative models in recommender systems. However, only a subsection (Section 2.4 in \cite{deldjoo2024review}) is dedicated to diffusion models with a few references, leaving much room to enhance the coverage.

On the other hand, several surveys with a focus on diffusion models have garnered the works in general methods and applications \cite{yang2023diffusion}, vision \cite{croitoru2023diffusion}, video \cite{xing2023survey}, medical imaging \cite{kazerouni2023diffusion}, and bioinformatics and computational biology \cite{guo2024diffusion}. Lin et al. \cite{lin2024survey} presented a survey on diffusion models in recommender systems, categorizing the role of diffusion models into three types: data engineering and encoding, recommendation model, and content presentation. At the intersection between diffusion models and recommender systems, this categorization is based on the perspective of diffusion models. Despite of its insights, we propose that after all, the recommendation task remains the important focus, as the purpose of adopting diffusion models is to enhance the recommendation performance, not vice versa: adapting the recommendation task to enable diffusion models. Therefore, we present an opposite and complementary perspective, focusing on the categorization of recommendation tasks. 

In this survey, we organize the literature along three orthogonal axes to provide a taxonomy of diffusion models in recommendation systems, as shown in Fig. \ref{fig:taxonomy_overview}.

\begin{enumerate}
    \item \textbf{Core Recommendation Tasks} focus on the fundamental prediction problems, including collaborative filtering and sequential recommendation. Within collaborative filtering, we examine settings characterized by different forms of auxiliary information, such as implicit feedback, explicit ratings, item graphs, and user graphs. For sequential recommendation, we treat point-of-interest (POI) recommendation as a representative special case and further categorize existing methods based on the role of historical sequences in diffusion models: (i) sequence as both diffusion target and guidance, (ii) sequence as diffusion target, and (iii) sequence as diffusion guidance.

    \item \textbf{Data Modality and Domain} capture the environments and data characteristics under which recommendation models operate. This axis encompasses image generation, multimodal recommendation, cross-domain recommendation, and text-to-recommendation (text-to-rec).

    \item \textbf{Trustworthy Objectives and Constraints} highlight the broader goals and practical considerations that shape modern recommender systems. We categorize existing work into four major aspects: fairness, accountability, transparency, and out-of-distribution (OOD).
    
\end{enumerate} 

\begin{figure}[t]
\centering
\begin{tikzpicture}[
    box/.style={
        draw,
        rounded corners,
        align=center,
        text width=0.85\linewidth,
        font=\small,
        inner sep=4pt
    }
]

\node[box, fill=gray!15] (title)
{\textbf{Diffusion Models in Recommendation Systems}};

\node[box, fill=red!15, below=3mm of title] (axis1)
{\textbf{Axis 1: Core Recommendation Tasks (What)}\\
Collaborative Filtering \quad | \quad Sequential Recommendation};

\node[box, fill=green!15, below=3mm of axis1] (axis2)
{\textbf{Axis 2: Data Modality and Domain (Where)}\\
Image Generation \quad | \quad Multimodal \quad | \quad Cross-Domain \quad | \quad Text-to-recommendation};

\node[box, fill=blue!15, below=3mm of axis2] (axis3)
{\textbf{Axis 3: Trustworthy Objectives and Constraints (Why)}\\
Fairness \quad | \quad Accountability \quad | \quad Transparency \quad | \quad Out-of-distribution };

\end{tikzpicture}
\caption{Three-axis taxonomy of diffusion models in recommendation systems.}
\label{fig:taxonomy_overview}
\end{figure}

By structuring the survey along these three axes, readers can more easily locate the research directions most relevant to their interests and understand how diffusion models are applied across tasks, contexts, and objectives. Additionally, we provide a dedicated section summarizing commonly used datasets and their properties, improving the comprehensiveness of the survey and supporting future research in this rapidly evolving field.

We organize the rest of the paper as follows. Section \ref{sec:foundation} introduces the foundation of diffusion models, Sections \ref{sec:core}, \ref{sec:data}, and \ref{sec:trustworthy} categorize the relevant papers based on the three axes, Section \ref{sec:datasets} discusses the datasets used by the relevant papers, Section \ref{sec:open} reveals open research directions to encourage future efforts, and Section \ref{sec:conclusion} concludes the survey paper.

\section{Foundation}
\label{sec:foundation}
In this section, we present the foundation of diffusion models and discuss subsequent advancements that enable their adoption in recommender systems, as shown in Fig. \ref{fig:overview}.

\subsection{Diffusion Models} \label{sec:diffusion}
\begin{figure}[t]
     \centering
     \begin{subfigure}[t]{0.48\textwidth}
         \centering
         \includegraphics[scale=0.25]{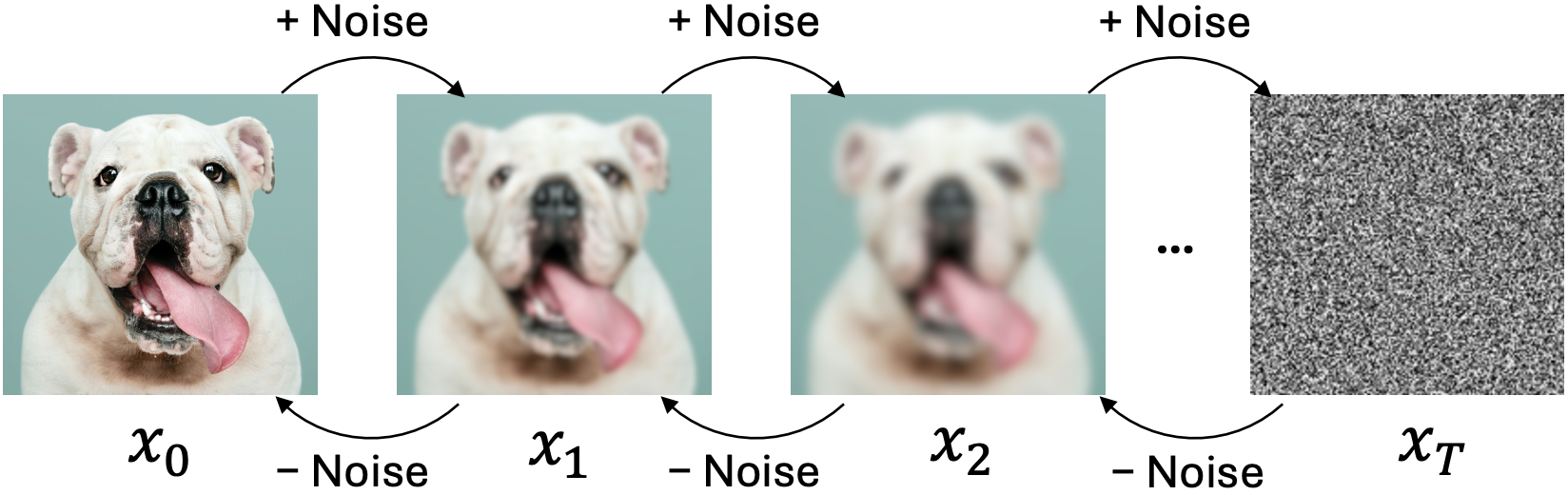}
         \caption{Illustration of the Diffusion Model. The forward process corrupts the data from left to right, followed by the reverse process for reconstruction.}
          \label{fig:diffusion_a}
     \end{subfigure}
     \hfill
     \begin{subfigure}[t]{0.48\textwidth}
         \centering
         \raisebox{3.5mm}{\includegraphics[scale=0.25]{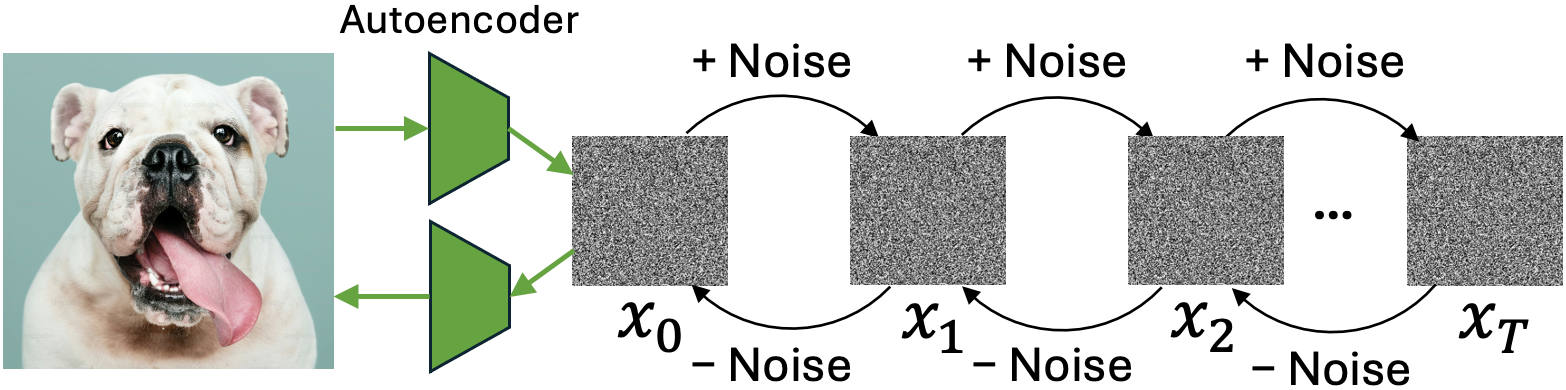}}
         \caption{Diffusion Model in Latent Space. The forward and reverse processes operate in the autoencoder’s latent space instead of the pixel space.}
          \label{fig:diffusion_b}
     \end{subfigure}
        \caption{Diffusion Models in Pixel Space and Latent Space.}
        \Description{Side by side plot of three chains of states.}
        \label{fig:diffusion}
\end{figure}

Generally, generative models can be categorized into two groups: probabilistic generative models that explicitly leverage probability distributions and generative adversarial networks that utilize adversarial learning. Probabilistic models usually are constrained to have a tractable normalizing constant to compute the likelihood or are required to approximate maximum likelihood training through surrogate objectives \cite{song2019generative}. Generative adversarial networks also have their limitations: instability in the training process \cite{salimans2016improved} and mode collapse, a situation where the model generates similar outputs with limited diversity \cite{metz2016unrolled}. As another branch of probabilistic generative models without these limitations, diffusion models operate under the general concept to model the gradient of the logarithmic probability density function of the data, also known as the score function \cite{liu2016kernelized}. These models do not require a tractable normalizing constant and can be trained with score-matching \cite{hyvarinen2005estimation}. An illustration of the diffusion model in discrete time steps is shown in Fig. \ref{fig:diffusion_a}. Starting with the initial image, the forward process adds noise to obtain a noisy image at the final time step, followed by the reverse process that removes noise to reconstruct the initial image. The utilization of diffusion models in recommender systems can be categorized into two streams: NCSN \cite{song2019generative} and DDPM \cite{ho2020denoising}. We discuss the two most original frameworks following their mathematical notation and subsequent improvements.

\begin{figure}
     \centering
     \begin{subfigure}[b]{0.3\textwidth}
         \centering
            \begin{tikzpicture}[scale=0.6, sibling distance=10em,
          every node/.style = {shape=rectangle, rounded corners,
            draw, align=center, font=\footnotesize,
            top color=white}]]
          \node [bottom color=red!40] {Diffusion Models}
            child [bottom color=red!20]{ node {NCSN \cite{song2019generative}} }
            child [bottom color=red!20]{ node {DDPM \cite{ho2020denoising}}
              child [bottom color=red!20, text width=4.5em]{ node {IDDPM \cite{nichol2021improved}}}};
        \end{tikzpicture}
         \caption{Diffusion Model Frameworks}
         \label{fig:overviewA}
     \end{subfigure}
     \hfill
     \begin{subfigure}[b]{0.3\textwidth}
         \centering
         \begin{tikzpicture}[scale=0.6, sibling distance=8em,
          every node/.style = {shape=rectangle, rounded corners,
            draw, align=center, font=\footnotesize,
            top color=white}]]
          \node [bottom color=green!40]{Efficient Enhancement}
            child [bottom color=green!20]{ node at (0, 0.25) {DDIM \cite{song2020denoising}} }
            child [bottom color=green!20]{ node at (0, -1.5) {Latent Diffusion Model \cite{rombach2022high}} }
            child [bottom color=green!20]{ node at (0, 0.25) {ODE Solver \cite{song2020score}} };
        \end{tikzpicture}
         \caption{Efficiency Enhancement}
         \label{fig:overviewB}
     \end{subfigure}
     \hfill
     \begin{subfigure}[b]{0.3\textwidth}
         \centering
         \begin{tikzpicture}[scale=0.6, sibling distance=10em,
          every node/.style = {shape=rectangle, rounded corners,
            draw, align=center, font=\footnotesize,
            top color=white}]]
          \node [bottom color=cyan!40]{Conditional Generation}
            child [bottom color=cyan!20]{ node at (0, -1) {Classifier \\ Guidance \cite{dhariwal2021diffusion}} }
            child [bottom color=cyan!20]{ node at (0, -1) {Classifier-free \\ Guidance \cite{ho2022classifier}} };
        \end{tikzpicture}
         \caption{Conditional Generation}
         \label{fig:overviewC}
     \end{subfigure}
        \caption{Structure of the Foundation. We structure the foundation of diffusion models in recommender systems into three topics: diffusion model frameworks (Sections \ref{sec:diffusion} and \ref{sec:improvement}), efficiency enhancement (Section \ref{sec:efficiency}), and conditional generation (Section \ref{sec:conditional}).}
        \Description{tree structure displaying topics}
        \label{fig:overview}
\end{figure}


\textbf{NCSN.} Song et al. \cite{song2019generative} introduce a framework of diffusion models that estimate the score function using score-matching and sample with Langevin dynamics. Given a dataset containing samples $\{x_1, x_2, \ldots, x_n\}$ from the underlying data distribution $p(x)$, a neural network aims to model the score function $\nabla_{x} \log p(x)$ to generate new samples for $p(x)$. Since gradients are hard to estimate on low-dimensional manifolds, they add to the data different levels of Gaussian noise $q_\sigma(\tilde{x}|x) = N(\tilde{x}|x, \sigma_i^2I)$, where $\{\sigma_i\}_{i=1}^L$ represents a positive geometric sequence of terms with length $L$. The noise levels are carefully chosen so that the first one is large enough and the last one is small enough to control its effect on the data. The noise conditional score network (NCSN) $s_\theta(x, \sigma)$ is trained to jointly estimate the scores at all noise levels, with the objective function as:

\begin{equation}
    \label{eqn:ncsn_losses}
    \mathcal{L}_{NCSN}(\theta; \{\sigma_i\}_{i=1}^L) =  \frac{1}{2L} \mathbb{E}_{p(x), \tilde{x}} \sum_{i=1}^L \lambda(\sigma_i) \left[ \norm{s_\theta(\tilde{x}, \sigma_i) + \frac{\tilde{x} - x}{\sigma_i^2}}^2_2 \right]
\end{equation}

\noindent where $\lambda(\sigma_i)$ is a positive coefficient function chosen to be $\sigma_i^2$ by the authors from empirical results. The objective function is minimized when $s_\theta(x, \sigma_i) = \nabla_x \log q_\sigma(x)$ for all $i \in \{1,2,\dots,L\}$.
As first introduced in the image domain, the output of the network has the same dimension as its input, so the model leverages the architecture of a U-net \cite{ronneberger2015u}. Additionally, dilated and atrous convolution are incorporated for their strengths \cite{chen2017deeplab, yu2016multi, yu2017dilated}, and conditional instance normalization \cite{dumoulin2016learned} helps incorporate the noise level at which the network is estimating for.

The inference relies on sampling with annealed Langevin dynamics, inspired by simulated annealing \cite{kirkpatrick1983optimization} and annealed importance sampling \cite{neal2001annealed}. The sampling algorithm starts with random noise and runs Langevin dynamics sequentially at each noise level. The first iteration samples from $q_{\sigma_1}(x)$ with step size $\delta_1$, and the second iteration of Langevin dynamics starts from the final sample of the last iteration with a reduced step size. This repeats for all noise levels, and the final sample of the last noise level is the final generation.


\textbf{DDPM.} As the other diffusion model framework, DDPM \cite{ho2020denoising} run a forward process to add Gaussian noise to the input, followed by a reverse process to denoise and reconstruct the original input in an iterative manner. Leveraging previous work \cite{sohl2015deep}, diffusion models follow a Markov chain, a sequence of states where the current state only depends on the previous state and no other states. Given the input $x_0 \sim q(x_0)$ and subsequent states $x_1, \dots, x_T$, the forward process with known Gaussian transitions is defined as $q(x_t|x_{t-1})=\mathcal{N}(x_t;\sqrt{1-\beta_t}x_{t-1}, \beta_tI)$, where $\beta_1, \dots, \beta_T$ represents a noise schedule which determines the scale of noise added to each step in the Markov chain and was first designed to be a linear schedule starting from $\beta_1=0.0001$ to $\beta_T=0.02$ over $T=1,000$ steps. Instead of computing the transition for $T$ times from $x_0$ to $x_{T}$, $x_T$ can be obtained in one step as $q(x_t | x_0) = \mathcal{N}\left(x_t; \sqrt{\bar{\alpha}_t} x_0, (1 - \bar{\alpha}_t) I\right)$, where $\alpha_t = 1 - \beta_t$ and $\bar{\alpha}_t = \prod_{s=1}^{t} \alpha_s$. The reverse process leverages a neural network parameterized by $\theta$ to learn the reverse transitions $p_\theta(x_{t-1}|x_t)=\mathcal{N}(x_{t-1};\mu_\theta(x_t,t), \Sigma_t=\sigma_t^2I)$, where $\mu_\theta(x_t,t)$ is the mean estimated by the network. After conditioning on $x_0$, the forward process posterior $q(x_{t-1} | x_t, x_0) = \mathcal{N}\left(x_{t-1}; \tilde{\mu}_t(x_t, x_0), \tilde{\beta}_t I\right)$ becomes tractable and a suitable target for learning the reverse transitions $p_\theta(x_{t-1}|x_t)$ with KL-divergence, resulting in the training objective as:





\begin{equation}
    \label{eqn:ddpm_x}
    \mathcal{L}_{DDPM-x}(\theta; \{\beta_i\}_{i=1}^T) =  E_{p(x), t, \epsilon} \left[ \frac{1}{2\sigma_t^2} \| \tilde{\mu}_t(x_t, x_0) - \mu_\theta(x_t, t) \|^2 \right]
\end{equation}

\noindent where $\tilde{\mu}_t(x_t, x_0) = \frac{\sqrt{\bar{\alpha}_{t-1} \beta_t}}{1 - \bar{\alpha}_t} x_0 + \frac{\sqrt{\alpha_t (1 - \bar{\alpha}_{t-1})}}{1 - \bar{\alpha}_t} x_t$ and $\mu_\theta(x_t, t) = \frac{\sqrt{\bar{\alpha}_{t-1} \beta_t}}{1 - \bar{\alpha}_t} \hat{x}_0 + \frac{\sqrt{\alpha_t (1 - \bar{\alpha}_{t-1})}}{1 - \bar{\alpha}_t} x_t$. The training is denoted as $x$-prediction due to its objective of estimating the mean of the posterior distribution. By replacing $x_t$ with a form of $\epsilon$, i.e. $x_t(x_0, \epsilon) = \sqrt{\bar{\alpha}_t} x_0 + \sqrt{1 - \bar{\alpha}_t} \epsilon$, the $\epsilon$-prediction training objective follows:

\begin{equation}
    \label{eqn:ddpm_e}
    \mathcal{L}_{DDPM-\epsilon}(\theta; \{\beta_i\}_{i=1}^T) = E_{p(x), t, \epsilon} \left[ \frac{\beta_t^2}{2\sigma_t^2\alpha_t (1 - \bar{\alpha}_t)} \| \epsilon-\epsilon_\theta( \sqrt{\bar{\alpha}_t} x_0 + \sqrt{1 - \bar{\alpha}_t} \epsilon, t )\|^2 \right]
\end{equation}

\noindent DDPM \cite{ho2020denoising} drops the coefficient in the training objective due to its negligible effect on the model performance. The training and inference algorithms can be found in Algorithms \ref{alg:ddpm_t} and \ref{alg:ddpm_i}. 

\begin{algorithm}
\caption{DDPM Training}
\label{alg:ddpm_t}
\begin{algorithmic}
\Require \(\{\beta_i\}_{i=1}^T\), \(T\)
\State \textbf{repeat}
\State $x_0 \sim q(x_0)$
\State $t \sim \text{Uniform}(\{1, \ldots, T\})$
\State $\epsilon \sim \mathcal{N}(0, I)$
\State Take gradient descent step on $\nabla_\theta \| \epsilon - \epsilon_\theta( \sqrt{\bar{\alpha}_t} x_0 + \sqrt{1 - \bar{\alpha}_t} \epsilon, t )\|^2$
\State \textbf{until} converged
\State
\Return \(\epsilon_\theta\)
\end{algorithmic}
\end{algorithm}

\begin{algorithm}
\caption{DDPM Inference}
\label{alg:ddpm_i}
\begin{algorithmic}
\Require \(T\), \(\{\beta_i\}_{i=1}^T\), \(\{\sigma_i\}_{i=1}^T\), \(\epsilon_\theta\)
\State $x_T \sim \mathcal{N}(0, I)$
    \For{$t = T, \ldots, 1$}
        \State $z \sim \mathcal{N}(0, I)$ if $t>1$, else $z=0$
        \State $x_{t-1} = \frac{1}{\sqrt{\alpha_t}} \left( x_t - \frac{1 - \alpha_t}{\sqrt{1 - \bar{\alpha}_t}} \epsilon_\theta(x_t, t) \right) + \sigma_t z$
    \EndFor
\State \Return $x_0$
\end{algorithmic}
\end{algorithm}

\subsection{Improvements to the Original Framework} \label{sec:improvement}
\textbf{Upgrades to DDPM.} Improvements \cite{nichol2021improved} are introduced to advance the DDPM framework \cite{ho2020denoising}, including a cosine noise schedule, importance sampling, and learning the variance $\Sigma_\theta(x_t, t)$. Compared to the linear schedule, the cosine schedule achieves better sample quality by having a more steady decline in the middle with minor changes near both ends to avoid large changes in noise levels:

\begin{equation}
    \label{eqn:cosine}
    \beta_t = 1-\frac{\bar{\alpha}_t}{\bar{\alpha}_{t-1}}, \quad \bar{\alpha}_t = \frac{f(t)}{f(0)}, \quad f(t) = \cos\left( \frac{t/T+s}{1+s}\cdot\frac{\pi}{2} \right)^2
\end{equation}

\noindent where $s$ is a small offset to prevent $\beta_t$ from being too small near $t=0$.

Due to the high variance across diffusion steps, applying importance sampling over the time step $t$ with weights $w_t \propto \sqrt{ E[L^2_t]}$ helps stabilize training, where $E[L^2_t]$ is estimated using the 10 most recent loss values recorded for each time step. At the start of training, $t$ is sampled uniformly until 10 samples have been drawn for every time step.

Although DDPM \cite{ho2020denoising} model $\mu_\theta(x_t, t)$ with fixed variance $\Sigma_t$, modeling both terms together achieves higher log-likelihood and requires fewer sampling steps. $\Sigma_\theta(x_t, t)$ is constructed as the interpolation between $\beta_t$ and $\tilde{\beta}_t$: $\Sigma_\theta(x_t, t) = \exp\left(v \log \beta_t + (1 - v) \log \tilde{\beta}_t \right)$, where $v$ is a vector output by the neural network. 

\subsection{Efficiency Enhancement} \label{sec:efficiency}
\begin{figure}
     \centering
     \begin{subfigure}{0.45\textwidth}
         \centering
         \includegraphics[scale=0.4]{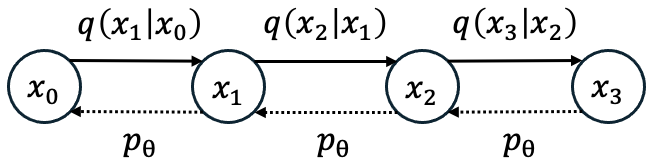}
         \caption{Markovian Processes in DDPM \cite{ho2020denoising}}
     \end{subfigure}
     \hfill
     \begin{subfigure}{0.45\textwidth}
         \centering
         \includegraphics[scale=0.4]{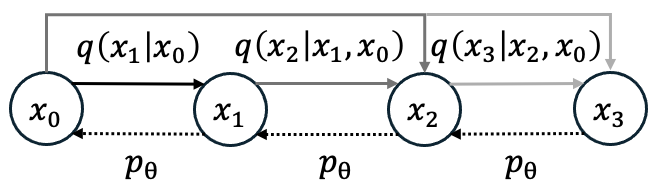}
         \caption{Non-Markovian Processes in DDIM \cite{song2020denoising}}
     \end{subfigure}
        \caption{Process Structure in DDPM \cite{ho2020denoising} and DDIM \cite{song2020denoising}}
        \Description{Side by side plot of two chains of states.}
        \label{fig:ddpmddim}
\end{figure}

\textbf{DDIM.} Meanwhile, Denoising Diffusion Implicit Models (DDIM) \cite{song2020denoising} improve upon DDPM \cite{ho2020denoising} by making sampling more efficient through formulating the diffusion processes as non-Markovian (Fig. \ref{fig:ddpmddim}a) instead of Markovian in DDPM \cite{ho2020denoising} (Fig. \ref{fig:ddpmddim}b). The difference lies in the extra condition on $x_0$ for each state, and the forward process posterior becomes $q_\gamma(x_{t-1} | x_t, x_0) = \mathcal{N} \left( \sqrt{\bar{\alpha}_{t-1}} x_0 + \sqrt{1 - \bar{\alpha}_{t-1} - \gamma_t^2} \cdot \left(\frac{x_t-\sqrt{\bar{\alpha}_t} x_0}{\sqrt{1 - \bar{\alpha}_t}} \right), \gamma_t^2 I \right)$, where $\gamma$ controls the stochasticity of the forward process. The training objective is equivalent to that in DDPM \cite{ho2020denoising}. Setting $\gamma_t$ to zero accelerates the sampling process by making those intermediate states $x_t$ deterministic as an implicit probabilistic model \cite{mohamed2016learning}, thereby sparing their computation. This achieves better sample quality than DDPM \cite{ho2020denoising} at fewer time steps, allowing the tradeoff between sample quality and efficiency. As a special case when $\gamma_t = \sqrt{\frac{1 - \bar{\alpha}_{t-1}}{1 - \bar{\alpha}_t}} \sqrt{1-\frac{\bar{\alpha}_t}{\bar{\alpha}_{t-1}}}$, the process becomes Markovian as DDPM \cite{ho2020denoising}.

\textbf{Efficient Sampling via ODE Solvers.} In parallel, Song et al. \cite{song2020score} propose a general framework of diffusion models from the aspect of stochastic differential equations (SDE) whose reverse process can be solved by numerical approaches to enhance efficiency. The SDE corresponds to a deterministic process with the same marginal probability densities, which satisfies to be an ordinary differential equation (ODE) that is more efficiently computed by existing solvers. 



\noindent NCSN and DDPM can be viewed as discrete-time approximations of two different types of continuous diffusion processes under the SDE framework. Specifically, NCSN corresponds to the Variance Exploding (VE) SDE, where the noise variance increases over time, while DDPM corresponds to the Variance Preserving (VP) SDE, where the total variance remains roughly constant throughout the diffusion process.


\textbf{Latent Diffusion Model.} Applying the diffusion process in the latent space can further improve efficiency. Rombach et al. \cite{rombach2022high} perform diffusion in the latent space instead of the pixel space, improving computational efficiency from a different perspective. As shown in Fig. \ref{fig:diffusion_b}, an encoder first maps samples into the latent space for the diffusion process, and a decoder subsequently reconstructs the samples from the latent representation. This requires the pre-training of an autoencoder, but significantly lowers the computational cost of latent diffusion models. 


\subsection{Conditional Generation} \label{sec:conditional}
To generate samples according to a condition $y$ (such as a class label or text prompt), diffusion models rely on classifier \cite{dhariwal2021diffusion} and classifier-free \cite{ho2022classifier} guidance. Classifier guidance \cite{dhariwal2021diffusion} incorporates a classifier, parameterized by $\psi$, in the reverse process: $p_{\theta, \psi}(x_{t-1} | x_t, y) \propto p_{\theta}(x_{t-1} | x_t) p_{\psi}(y | x_{t-1})$. Taylor expansion enables sampling from a Gaussian distribution: $\mathcal{N}(\mu_\theta(x_t,t)+g \Sigma_\theta(x_t,t) \nabla_{x_{t-1}} \log p_\psi(y | x_{t-1}), \Sigma_\theta(x_t,t))$, where $g$ represents the gradient scale. Samples are generated towards where the class label has a higher likelihood by shifting the mean according to the classifier.


To alleviate the need of a separate classifier, classifier-free guidance \cite{ho2022classifier} jointly trains a conditional and an unconditional denoising model through a single neural network: $\epsilon_\theta(x_t, t, y) = (1 + g) \epsilon_\theta(x_t, t, y) - g \epsilon_\theta(x_t, t, \emptyset)$. The training algorithm randomly switches between the conditional and unconditional frameworks with probability $p_{rand}$, either providing the class representation or setting it to null ($y=\emptyset$).

Armed with the foundation described above on score-based generative models (NCSN \cite{song2019generative} and DDPM \cite{ho2020denoising}) through $x$-prediction, $\epsilon$-prediction, further improvements on performance \cite{nichol2021improved} and efficiency \cite{rombach2022high, song2020denoising, song2020score}, and conditional generation \cite{ho2022classifier, dhariwal2021diffusion}, we dive into the application of diffusion models in recommendation problems in the next section.

\section{Core Recommendation Tasks}
\label{sec:core}
Diffusion models have been applied to multiple core recommendation settings, most prominently collaborative filtering and sequential recommendation. Although these tasks differ in their modeling assumptions and input structure, diffusion models are typically incorporated through a small number of practical paradigms. In general, diffusion-based recommendation systems operate in one of two ways:

\begin{figure}[t]
\centering
\begin{tikzpicture}[
    edge from parent/.style={draw, -latex}, 
    every node/.style={draw, rounded corners, align=center, font=\small},
    level 1/.style={level distance=0.8cm, sibling distance=8cm}, 
    level 2/.style={level distance=0.8cm, sibling distance=3.5cm},   
    grow=down
]

\node[fill=blue!15] {Core Recommendation Tasks}
    child { node[fill=orange!20] {Data Augmentation}
        child { node[fill=orange!10] {Add synthetic interactions} child{ node[fill=orange!5] {Train recommender}}} 
        child { node[fill=orange!10] {Generate negatives} child{ node[fill=orange!5] { Contrastive learning}}}
    }
    child { node[fill=green!20] {Direct Recommendation}
        child { node[fill=green!10] {Denoise interaction vectors} child{ node[fill=green!5] {Rank by probability}}}
        child { node[fill=green!10] {Denoise embeddings} child{ node[fill=green!5] {Rank by similarity}}}
    };
\end{tikzpicture}
\caption{Practical Usage of Diffusion Models in Recommender Systems on Core Recommendation Tasks. For data augmentation, the diffusion model can add synthetic interactions to strengthen the downstream recommender model or generate negative samples to facilitate contrastive learning. For direct recommendation, the diffusion model can denoise interaction vectors and rank by the denoised probabilties or denoise the user/item embeddings and rank by similarity against existing items.}
\Description{Structure of topics}
\label{fig:diffusion_practical}
\end{figure}

\begin{figure}[t]
     \centering
     \begin{subfigure}{0.32\textwidth}
         \centering
         \includegraphics[scale=0.3]{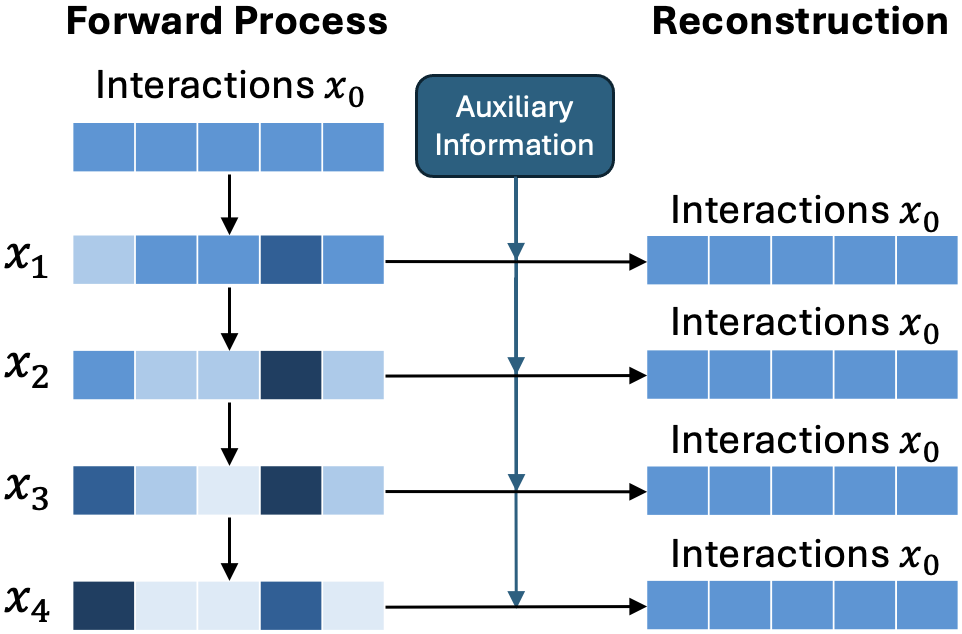}
         \caption{Training Process to Reconstruct}
     \end{subfigure}
     \hspace{0.035\textwidth}
     \begin{subfigure}{0.32\textwidth}
         \centering
         \includegraphics[scale=0.3]{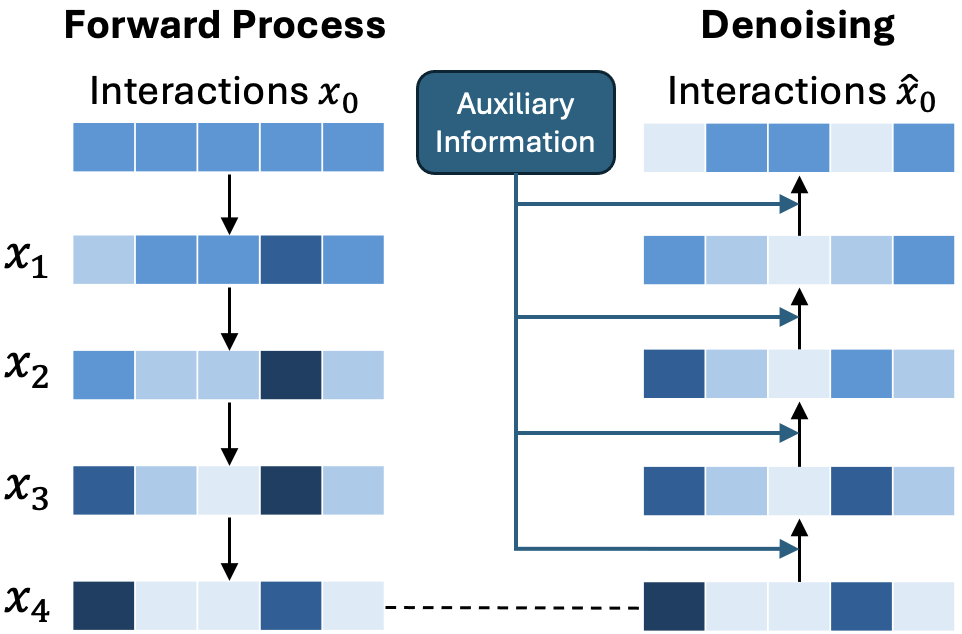}
         \caption{Inference Process to Denoise}
     \end{subfigure}
     \hspace{0.01\textwidth}
     \begin{subfigure}{0.28\textwidth}
         \centering
         \includegraphics[scale=0.28]{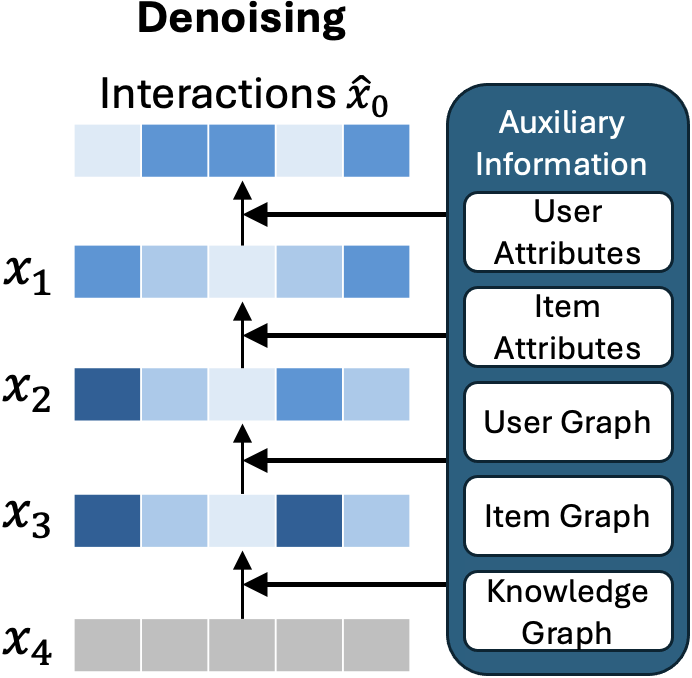}
         \caption{Inference Process to Generate}
     \end{subfigure}
        \caption{Training and Inference of Diffusion Models with User Interaction Vectors. Training adds noise and reconstructs the original interaction vector with auxiliary information. Inference denoises the original vector or generates the vector from noise using auxiliary information which includes user/item attributes, and user/item/knowledge graphs.}
        \Description{Side by side plot of three chains of states.}
        \label{fig:cf}
\end{figure}

\begin{enumerate}
    \item \textbf{Data augmentation:} The diffusion model serves as a generative module that enhances a downstream recommender. It may generate synthetic user-item interactions to densify sparse data or produce high-quality negative samples to facilitate contrastive learning and robust training.

    \item \textbf{Direct generative recommendation:} The diffusion model functions as the core recommender. It learns to denoise corrupted interaction vectors or embeddings and produces recommendations either by ranking reconstructed interaction probabilities or by measuring similarity between denoised user/item representations and candidate items.
\end{enumerate}

Fig.~\ref{fig:diffusion_practical} summarizes these practical uses of diffusion models in recommender systems. This paradigm-level distinction applies across different recommendation tasks and provides a unifying perspective before discussing task-specific instantiations.

\subsection{Collaborative Filtering}
Collaborative filtering remains a core task in recommendation, which considers the preferences of other users to recommend items to a user. Fig. \ref{fig:cf} illustrates how diffusion models generally operate in collaborative filtering. During training, noise is added to the user interaction vector, and the model learns to reconstruct the original vector, often leveraging auxiliary information. At inference, the model either denoises a partially corrupted interaction vector or generates it from noise, conditioning on auxiliary information. 

Table~\ref{tab:empirical_cf} presents an empirical comparison of representative diffusion-based collaborative filtering models on three datasets. Diffusion approaches consistently outperform the strongest baselines reported in their respective papers on both Recall@20 and NDCG@20. Based on the type of feedback and auxiliary information, we categorize the papers into implicit feedback, explicit ratings, item graph, user graph, and knowledge graph, as shown in Fig. \ref{fig:taxonomycore}.

\begin{table}[t]
\centering
\small
\caption{Empirical comparison of representative diffusion-based recommender models from top-tier venues in collaborative filtering. The strongest baseline reported in the original papers is chosen for comparison, and the relative gain (\% Gain) indicates the improvement over this baseline. All reported metrics are taken directly from the respective publications.}
\label{tab:empirical_cf}
\begin{tabular}{l|c|ccc|ccc}
\toprule
Dataset & Model & Recall@20 & Baseline & \% Gain & NDCG@20 & Baseline & \% Gain \\
\midrule
\multirow{4}{*}{\shortstack[l]{Amazon-Books \\ 
\cite{mcauley2013hidden,he2016ups,mcauley2015image,ni2019justifying,hou2024bridging}}} & HDRM \cite{yuan2025hyperbolic}& 0.1057 & DiffRec \cite{wang2023diffusion}& +4.7\% & 0.0582 & DiffRec \cite{wang2023diffusion}& +6.5\% \\
 & DiffRec \cite{wang2023diffusion}& 0.1010 & MultiVAE \cite{liang2018variational}& +8.02\% & 0.0547 & MultiVAE \cite{liang2018variational}& +12.78\% \\
 & DDRM \cite{zhao2024denoising}& 0.0813 & BOD \cite{wang2023efficient} & +1.50\% & 0.0396 & BOD \cite{wang2023efficient} & +2.86\% \\
  & BSPM \cite{choi2023blurring}& 0.0733 & LinkProp \cite{fu2022revisiting} & +1.66\% & 0.0610 & LinkProp \cite{fu2022revisiting} & +3.74\% \\
\midrule
\multirow{4}{*}{MovieLens-1M \cite{harper2015movielens}} & HDRM \cite{yuan2025hyperbolic}& 0.1852 & DiffRec \cite{wang2023diffusion}& +4.2\% & 0.1190 & DiffRec \cite{wang2023diffusion}& +4.8\% \\
 & CF-Diff \cite{hou2024collaborative}& 0.1843 & DiffRec \cite{wang2023diffusion}&  +4.54\% & 0.1176 & DiffRec \cite{wang2023diffusion}& +3.98\% \\
 & DiffRec \cite{wang2023diffusion}& 0.1787 & MultiDAE++ \cite{liang2018variational} & +0.90\% & 0.1148 & CODIGEM \cite{walker2022recommendation} & +6.69\% \\
  & DDRM \cite{zhao2024denoising}& 0.1261 & BOD \cite{wang2023efficient} & +2.19\% & 0.0739 & DeCA \cite{wang2022learning} & +9.81\% \\
\midrule
 \multirow{5}{*}{Yelp} & HDRM \cite{yuan2025hyperbolic}& 0.1024 & HICF \cite{yang2022hicf} & +5.8\% & 0.0499 & HICF \cite{yang2022hicf} & +2.4\% \\
 & CF-Diff \cite{hou2024collaborative}& 0.0962 & SGL \cite{wu2021self} & +1.91\% & 0.0480 & DiffRec \cite{wang2023diffusion}& +6.19\% \\
 & DiffRec \cite{wang2023diffusion}& 0.0960 & MultiVAE \cite{liang2018variational}& +1.59\% & 0.0478 & MultiVAE \cite{liang2018variational}& +4.37\% \\
 & DDRM \cite{zhao2024denoising}& 0.0860 & T-CE \cite{wang2021denoising}& +2.38\% & 0.0415 & BOD \cite{wang2023efficient} & +2.98\% \\
 & BSPM \cite{choi2023blurring}& 0.0720 & SimpleX \cite{mao2021simplex} & +2.71\% & 0.0593 & MGDCF \cite{hu2024mgdcf} & +3.13\% \\

\bottomrule
\end{tabular}
\end{table}

\subsubsection{\textbf{Implicit Feedback}} Diffusion models are widely adopted to enhance recommendation under implicit feedback by reconstructing user-item interactions and refining learned representations. Early works directly denoise the user-item interaction vector, generating cleaner interaction signals for ranking and representation learning \cite{walker2022recommendation, wang2023diffusion, lee2025collaborative}. These methods are simple and easily integrated into existing frameworks but may be limited in handling sparsity or cold-start scenarios. Subsequent extensions further refine interaction reconstruction by selecting high-confidence edges \cite{xie2025diffmsr} or conditioning on similar users \cite{wen2026condiff}, improving robustness and personalization at the cost of added complexity. 

Diffusion has also been extended to heterogeneous interaction graphs, where multiple edge types are reconstructed in latent space or adjacency form \cite{li2025diffgraph, li2025diffusionkan}. Compared to simpler denoising approaches, these methods better capture complex relationships but require more sophisticated graph modeling. In many cases, denoised interactions are integrated into existing recommendation backbones, including LightGCN \cite{he2020lightgcn, li2025gddrec} and hyperbolic encoders \cite{chami2019hyperbolic, yuan2025hyperbolic}, or used within model-agnostic frameworks that enhance contrastive learning through reconstructed embeddings \cite{zhao2024denoising, hao2025diff, yang2024diffgcl, liu2025ggdhscl, jing2025interest}. These integrations demonstrate flexibility: backbone-based approaches leverage existing architectures efficiently, while model-agnostic frameworks provide more general enhancement at the cost of additional training. Diffusion models have further been leveraged to mitigate data imbalance by generating synthetic interactions \cite{taskin2024effects, kotelnikov2023tabddpm}.

Beyond direct reconstruction, many methods incorporate auxiliary signals to guide the reverse diffusion process. Conditioning on original interaction vectors, neighbor information, pretrained embeddings, or structural features enables more informative denoising and improved robustness \cite{buchanan2024incorporating, chen2025conditional, yi2024directional, peng2024diffusion, hou2024collaborative, zhang2025gdiffmae}. Guidance can also stem from user embeddings, user-bundle interactions, or reasoning signals extracted from LLMs, strengthening personalization and cold-start performance \cite{bui2025personalized, qu2025generative, lee2024stochastic, zhu2025addressing, han2025diffusion}. These methods trade off additional complexity and potential computational overhead for improved adaptability in heterogeneous recommendation scenarios.

Diffusion models have been actively explored for bundle recommendation, where bundle- or item-level representations are reconstructed and aligned with user preferences \cite{li2025divide, zhang2025modeling, chen2025dual}. Approaches that corrupt and denoise bundle embeddings improve ranking performance, with design differences affecting flexibility versus computational cost.

In multi-behavior recommendation settings, diffusion models are employed to align and reconstruct representations across different behavior types, enabling semantic consistency and cross-behavior knowledge transfer. Some approaches incorporate latent confounders or behavior-specific interaction frequencies to guide graph and embedding reconstruction, improving robustness under heterogeneous behavioral signals \cite{chen2025causal, zheng2025diffusion}. Others denoise auxiliary behavior graphs into a target behavior representation, explicitly enforcing semantic alignment across behaviors \cite{mo2026hierarchical}. These designs balance robustness against heterogeneity with modeling complexity. Diffusion models have also been applied to click-through rate (CTR) prediction \cite{ni2025hierdiffuse, zhang2025dgenctr}.


Finally, while most approaches adopt continuous Gaussian diffusion, alternative formulations have been proposed. Discrete diffusion models based on Bernoulli processes reconstruct interaction bits \cite{benedict2023recfusion}, graph edges \cite{ju2025diffgr}, or target items \cite{hu2025fading}, and hybrid models combine discrete edge diffusion with continuous node embedding diffusion \cite{zhang2025graph}. Deterministic diffusion processes formulated as ODE-based blurring and sharpening operations provide efficient, training-free alternatives and connect diffusion to classical graph filtering methods \cite{choi2023blurring, shen2021powerful}. Across these settings, diffusion models offer a flexible and powerful framework for denoising and generating implicit feedback, improving representation quality, robustness, and recommendation accuracy.

\begin{figure*}[t]
\centering
\begin{tikzpicture}[
    root/.style={draw, rounded corners, minimum width=12mm, minimum height=8mm, fill=gray!20, font=\bfseries\small, align=center},
    mid/.style={draw, rounded corners, minimum width=15mm, minimum height=3mm, font=\bfseries\small, align=center},
    cf/.style={mid, fill=red!40},
    sr/.style={mid, fill=blue!40},
    sub1/.style={draw, rounded corners, minimum width=70mm, minimum height=3mm, font=\footnotesize, align=left, text width=110mm, fill=#1},
    short/.style={draw, rounded corners, minimum width=70mm, minimum height=3mm, font=\footnotesize, align=left, text width=110mm, fill=#1},
    sub/.style={draw, rounded corners, minimum width=70mm, minimum height=3mm, font=\footnotesize, align=left, text width=105mm, fill=#1},
    arrow/.style={
    -{Latex},
    line width=0.7pt,
    draw=black!70
}
]

\node[root, rotate=90] (core) at (-3,1) {Core Recommendation Tasks};

\node[cf] (cf) at (-1,3) {Collaborative \\ Filtering};
\node[sr] (sr) at (-1,-2) {Sequential \\ Recommendation};

\node[sub1=red!20] (cf_imp) at (6,3) {\textbf{Implicit Feedback}: BSPM \cite{choi2023blurring} Buchanan et al. \cite{buchanan2024incorporating} CF-Diff \cite{hou2024collaborative} CODIGEM \cite{walker2022recommendation} DDRM \cite{zhao2024denoising} DiffGT \cite{yi2024directional} DiffRec \cite{wang2023diffusion} Recfusion \cite{benedict2023recfusion} SCONE \cite{lee2024stochastic} Taskin et al. \cite{taskin2024effects} GDDRec \cite{li2025gddrec} CDiff4Rec \cite{lee2025collaborative} HDRM \cite{yuan2025hyperbolic} C-DiffRec \cite{chen2025conditional} DeftRec \cite{qu2025generative} Han and Chun \cite{han2025diffusion} Diff-GNDCRec \cite{hao2025diff} DiffUP \cite{peng2024diffusion} GDMCF \cite{zhang2025graph} DiffGCL \cite{yang2024diffgcl} MoDiffE \cite{li2025divide} DiffGraph \cite{li2025diffgraph} DST-GKAN \cite{li2025diffusionkan} DiffGR \cite{ju2025diffgr} GGDHSCL \cite{liu2025ggdhscl} DisCo \cite{bui2025personalized} CSDM \cite{zhu2025addressing} DiffCL2 \cite{jing2025interest} RDiffBR \cite{zhang2025modeling} GDiffMAE \cite{zhang2025gdiffmae} DiffMSR \cite{xie2025diffmsr} HGCLD \cite{mo2026hierarchical} DiffCL2 \cite{jing2025interest} RDiffBR \cite{zhang2025modeling} GDiffMAE \cite{zhang2025gdiffmae} DiffMSR \cite{xie2025diffmsr} HGCLD \cite{mo2026hierarchical} PreferGrow \cite{hu2025fading} HierDiffuse \cite{ni2025hierdiffuse} DGenCTR \cite{zhang2025dgenctr} DCDL \cite{chen2025dual} ConDiff \cite{wen2026condiff} CVID \cite{chen2025causal} DiffHGCN \cite{zheng2025diffusion}};
\node[short=red!20] (cf_exp) at (6,1.5) {\textbf{Explicit Ratings}: DGRM \cite{jiangzhou2024dgrm} EDGE-Rec \cite{priyam2024edge} DPDM \cite{wang2025differentially}};
\node[sub1=red!20] (cf_item) at (6,0.9) {\textbf{Item Graph}: DICES \cite{dong2024dices} G-Diff \cite{chen2024g} GiffCF \cite{zhu2024graph} DCBR \cite{li2025disentangled}};
\node[sub1=red!20] (cf_user) at (6,0.1) {\textbf{User Graph}: DiffuSAR \cite{zang2025diffusion} RecDiff \cite{li2024recdiff} SGSR \cite{liu2024score} GDSSL \cite{li2024graph} GDSR \cite{li2025dual} ARD-SR \cite{sun2025model} F-WADM \cite{li2025feature} GDSR \cite{gao2025graph}};
\node[sub1=red!20] (cf_knowledge) at (6,-0.8) {\textbf{Knowledge Graph}: DiffKG \cite{jiang2024diffkg} RsDiff \cite{cui2025rsdiff} DKGM \cite{zhao2025dkgm} TCM-DiffPR \cite{zhang2025tcm_diffpr} KMDCL \cite{li2025mask} KDiffE \cite{mo2025knowledge} DAGR \cite{shi2025dagr}};
\node[sub=blue!20] (sr_poi) at (6,-1.7) {\textbf{POI}: DCPR \cite{long2024diffusion} Diff-DGMN \cite{zuo2024diff} Diff-POI \cite{qin2023diffusion} DSDRec \cite{wang2024dsdrec} HGDRec \cite{pan2025hgdrec} DMSDRec \cite{li2025dmsdrec} UB-Diff \cite{zuo2025bridging}};
\node[sub=blue!20] (sr_targetguid) at (6,-2.7) {\textbf{Sequence as Diffusion Target and Guidance}: CaDiRec \cite{cui2024context} DCRec \cite{huang2024dual} SDRec \cite{chen2024semanticaware} DMI \cite{le2025diffusion} ADRec \cite{chen2025unlocking} GlobalDiff \cite{luo2025enhancing} InDiRec \cite{qu2025intent} DiffuASR \cite{liu2023diffusion} Diff-MSIN \cite{cui2025diffusion} HorizonRec \cite{zha2025align} DMMD4SR \cite{lu2025dmmd4sr} M$^3$BSR \cite{cui2025multi} DAC4Rec \cite{chen2025energy}};
\node[sub=blue!20] (sr_target) at (6,-4.0) {\textbf{Sequence as Diffusion Target}: DCDR \cite{lin2024discrete} DDSR \cite{xie2024breaking} DIDER \cite{wu2024diffusion} Diff4Rec \cite{wu2023diff4rec} MISD \cite{li2024multi} PDRec \cite{ma2024plug} S2IDM \cite{ma2024seedrec} SdifRec \cite{xie2024bridging} CADSR \cite{you2024context} DiffGCA \cite{li2024diffgca} HR4SR \cite{zheng2025feature} EDiffuRec \cite{lee2024ediffurec} SimDiffRec \cite{choi2025similarity} DiffusionGS \cite{li2025diffusiongs} FDISR \cite{feng2025fusion} DiffuMIN \cite{lai2025modeling} MDSBR \cite{li2025mdsbr} MDCR \cite{liu2025multi} DiffSBR \cite{wang2026diffsbr} CDP \cite{zhao2025adaptive} BBDRec \cite{bai2025unconditional} Ge et al. \cite{ge2025time}};
\node[sub=blue!20] (sr_guid) at (6,-5.4) {\textbf{Sequence as Diffusion Guidance}: CDDRec \cite{wang2024conditional} DiffRec2 \cite{du2023sequential} DiffRIS \cite{niu2024diffusion} DiffuRec \cite{li2023diffurec} DimeRec \cite{li2024dimerec} DreamRec \cite{yang2024generate} iDreamRec \cite{hu2024generate} IDSRec \cite{niu2025implicit} LeadRec \cite{wang2024leadrec} PreferDiff \cite{liu2024preference} DiffuRecSys \cite{zolghadr2024generative} GCDR \cite{wu2025learning} T2Diff \cite{wang2025unleashing} DiQDiff \cite{mao2025distinguished} TDM \cite{mao2025addressing} DAE4Rec \cite{song2025enhancing} F-WADM \cite{li2025feature} GDSR \cite{gao2025graph} D-GenAtt \cite{liu2025generate} ADIGRec \cite{zhu2025adaptive} TA-Rec \cite{mao2025efficiency} PRISM \cite{ma2025bridging} FairGENRec \cite{liu2025alleviating} DiffDiv \cite{cai2025unleashing}};

\draw[arrow] (core.south) |- (cf.west);
\draw[arrow] (core.south) |- (sr.west);

\draw[arrow] (cf.east) |- (cf_imp.west);
\draw[arrow] (cf.east) |- (cf_exp.west);
\draw[arrow] (cf.east) |- (cf_item.west);
\draw[arrow] (cf.east) |- (cf_user.west);
\draw[arrow] (cf.east) |- (cf_knowledge.west);

\draw[arrow] (sr.east) |- (sr_poi.west);
\draw[arrow] (sr.east) |- (sr_targetguid.west);
\draw[arrow] (sr.east) |- (sr_target.west);
\draw[arrow] (sr.east) |- (sr_guid.west);

\end{tikzpicture}
\caption{Taxonomy for Core Recommendation Tasks, including Collaborative Filtering and Sequential Recommendation.}
\label{fig:taxonomycore}
\end{figure*}
\subsubsection{\textbf{Explicit Ratings}} Diffusion models can denoise or reconstruct explicit rating matrices to improve prediction accuracy and robustness. DGRM \cite{jiangzhou2024dgrm} combines diffusion with a generative adversarial network, reinforcing reconstruction but introducing additional training complexity. EDGE-Rec \cite{priyam2024edge, chen2023efficient} reconstructs rating patches using user and item features, offering a simpler, more modular approach at the cost of capturing less global structure. For privacy-preserving applications, DPDM \cite{wang2025differentially} perturbs the rating frequency matrix and relies on diffusion-based denoising to recover information, trading some accuracy for strong privacy guarantees. Overall, these methods illustrate different trade-offs: DGRM emphasizes robust reconstruction, EDGE-Rec prioritizes modularity and simplicity, and DPDM focuses on privacy.

\subsubsection{\textbf{Item Graph}} Diffusion models can also leverage item graphs to improve recommendation by denoising item relationships and guiding user-item interactions. One common approach is to directly denoise the item graph, producing cleaner item embeddings that reduce false connections and facilitate contrastive learning with user and interaction data \cite{dong2024dices, li2025disentangled}. This method is simple and effective for enhancing embeddings but may not fully capture temporal or contextual patterns.

Alternatively, item graphs can support the reverse diffusion process by informing the reconstruction of user interaction vectors, for instance, through graph networks that weigh more recent interactions more heavily \cite{chen2024g}. Item graphs can also guide anisotropic diffusion, where the noise applied to each interaction is adjusted based on item relationships, improving reconstruction quality while avoiding the corruption of important signals \cite{zhu2024graph}. Compared to direct graph denoising, these guided approaches better capture relational and temporal dependencies but involve higher computational and modeling complexity. Across these methods, integrating the item graph in the diffusion process consistently strengthens embedding learning and enhances recommendation performance.




\subsubsection{\textbf{User Graph}} Diffusion models can also leverage user graphs to enhance collaborative filtering, either by directly diffusing user embeddings or by incorporating the user graph in the reverse denoising process. In the former approach, user embeddings are refined through graph-based denoising, which can improve social recommendation by jointly optimizing reconstruction and recommendation objectives \cite{li2024recdiff, li2024graph, li2025dual, liu2024score}. This method is simple and effective for improving social recommendation, but may not fully exploit neighborhood structures.

Contrastive learning is often applied, comparing denoised user graphs against the original user graph to strengthen representation learning. Alternatively, the user graph can guide the reverse diffusion process, where noise is added to user embeddings and denoising is conditioned on social connections or neighbors’ interactions, improving embedding quality for downstream recommendation tasks \cite{zang2025diffusion, sun2025model, li2025feature, gao2025graph}. Compared to simple denoising, guided approaches better capture relational and contextual information, though they involve higher computational cost and more complex modeling. Across these strategies, diffusion-based denoising and augmentation of the user graph consistently enhance the accuracy of embeddings and the performance of social and collaborative recommendation models.

\subsubsection{\textbf{Knowledge Graph}} Knowledge graphs have become an effective way to enhance collaborative filtering by providing structured relational information about users and items. A common strategy is to denoise the graph with diffusion models to improve representation learning by producing cleaner embeddings and more reliable relational patterns \cite{jiang2024diffkg, cui2025rsdiff, zhao2025dkgm, shi2025dagr}. This straightforward denoising improves embedding quality with minimal architectural changes but may not fully exploit complex relational structures.

Corrupted or selectively augmented graphs can also be used in contrastive learning to better capture latent user-item preferences \cite{li2025mask}, or focus on high-confidence relations to generate enriched graphs \cite{zhang2025tcm_diffpr}. Compared to simple denoising, these approaches provide richer relational information and stronger embedding learning, but require additional computation and careful design to select or augment edges effectively. By integrating these denoised or enriched knowledge graphs into recommendation models, systems can learn more accurate embeddings and deliver higher-quality recommendations.






\begin{figure}[t]
     \centering
     \begin{subfigure}{0.32\textwidth}
         \centering
         \includegraphics[scale=0.4]{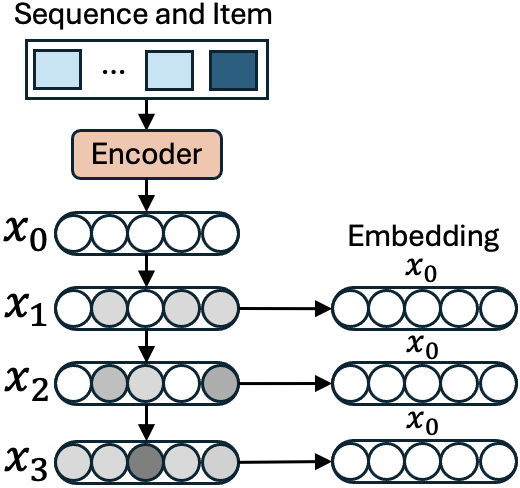}
         \caption{Sequence as Target}
     \end{subfigure}
     \hfill
     \begin{subfigure}{0.32\textwidth}
         \centering
         \includegraphics[scale=0.4]{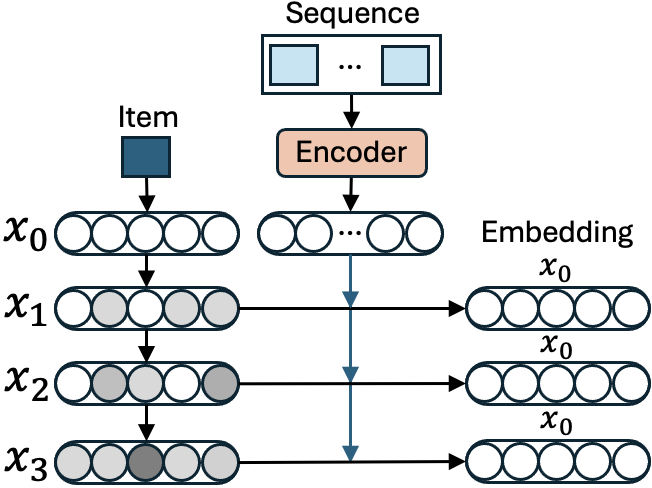}
         \caption{Sequence as Guidance}
     \end{subfigure}
     \hfill
     \begin{subfigure}{0.32\textwidth}
         \centering
         \includegraphics[scale=0.4]{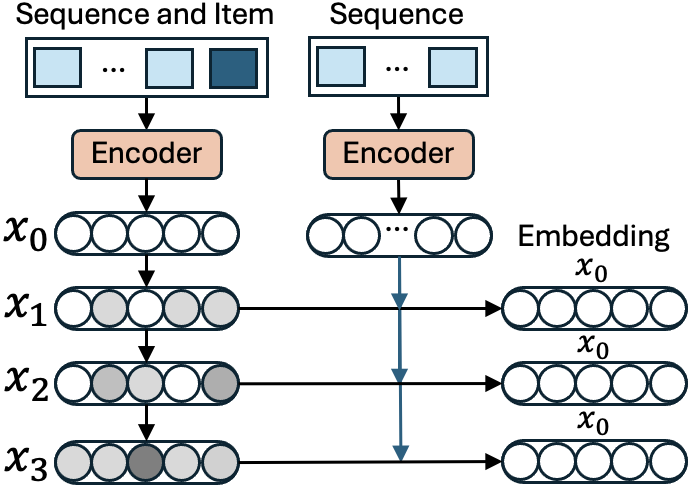}
         \caption{Sequence as Target and Guidance}
     \end{subfigure}
        \caption{Sequential Recommendation with Diffusion Models. The sequence can serve as the diffusion target, the diffusion guidance, or both. As the target, the sequence is reconstructed in the diffusion process. As guidance, the sequence guides the reverse process to recover the target item embedding. As the target and guidance, the sequence guides the reverse process to recover the sequence.}
        \Description{Side by side plot of three chains of states.}
        \label{fig:sequential}
\end{figure}

\subsection{Sequential Recommendation}
Sequential recommendation has received increasing attention in integrating diffusion models into recommender systems, where the goal is to predict the next item a user will interact with. Fig. \ref{fig:sequential} illustrates how diffusion models generally operate in sequential recommendation. When the sequence serves as the diffusion target, it is reconstructed through the diffusion process. When it serves as guidance, it informs the reverse process to recover the target item embedding. In cases where it functions as both target and guidance, the sequence guides the reverse process to reconstruct itself while informing the target item embeddings.

Table~\ref{tab:empirical_sr} compares representative diffusion-based sequential recommendation models on three datasets. Diffusion-based methods consistently outperform the strongest baselines reported in their respective papers in terms of HR@20 and NDCG@20. The early approach, DreamRec \cite{yang2024generate}, demonstrates substantial gains over contrastive models, and more recent variants enlarge the margin by refining guidance mechanisms. 

Besides recognizing POI recommendation as a special case of sequential recommendation in Fig. \ref{fig:taxonomycore}, we categorize other approaches based on their use of the historical sequence into three subgroups: sequence as diffusion target and guidance, sequence as diffusion target, and sequence as diffusion guidance. In these approaches, the historical sequence can serve as the subject of the diffusion process, as guidance in the reverse process, or both.

\subsubsection{\textbf{POI}} Diffusion models have been applied to POI recommendation to capture both sequential and spatial patterns in user trajectories. A common approach is to diffuse user or POI embeddings and reconstruct them using guidance from past trajectories, short-term mobility patterns, or global representations \cite{long2024diffusion, wang2024dsdrec, pan2025hgdrec}. These methods are straightforward and effective for modeling recent movement patterns but may be less capable of capturing global trajectory context. 

SDE formulations diffuse target location embeddings with the reverse process conditioned on user embeddings and optimized for both reconstruction and prediction \cite{qin2023diffusion, zuo2024diff}. Compared to basic embedding diffusion, SDE-based methods provide smoother and more flexible reconstruction, though at the cost of increased computational complexity.

Other methods denoise spatio-temporal latent representations \cite{li2025dmsdrec} or bridge target representations into user embeddings using auxiliary transition information \cite{zuo2025bridging, zhou2024denoising}. These methods emphasize capturing both sequential and spatial dependencies, offering richer trajectory modeling, but require careful design of input features or auxiliary signals. By integrating these denoised or generated embeddings, POI recommendation models can more accurately predict the next location a user is likely to visit.

\begin{table}[t]
\centering
\small
\caption{Empirical comparison of representative diffusion-based recommender models from top-tier conferences in sequential recommendation. The strongest baseline reported in the original papers is chosen for comparison, and the relative gain (\% Gain) indicates the improvement over this baseline. All reported metrics are taken directly from the respective publications.}
\label{tab:empirical_sr}
\begin{tabular}{l|c|ccc|ccc}
\toprule
Dataset & Model & HR@20 (\%) & Baseline & \% Gain & NDCG@20 (\%) & Baseline & \% Gain \\
\midrule
\multirow{5}{*}{YooChoose \cite{ben2015recsys}} & ADIGRec \cite{zhu2025adaptive} & 7.91 & DimeRec \cite{li2024dimerec} & +28.83\% & 3.05 & DimeRec \cite{li2024dimerec} & +32.03\% \\
& TDM \cite{mao2025addressing} & 6.90 & PDRec \cite{ma2024plug} & +9.85\% & 4.34 & PDRec \cite{ma2024plug} & +26.95\% \\
 & TA-Rec \cite{mao2025efficiency} & 6.15 & PreferDiff \cite{liu2024preference} & +7.14\% & 4.06 & PDRec \cite{ma2024plug} & +31.82\% \\
 & DreamRec \cite{yang2024generate} & 4.78 & CL4SRec \cite{xie2022contrastive} & +7.42\% & 2.23 & CL4SRec \cite{xie2022contrastive} & +19.89\% \\
& DiffDiv \cite{cai2025unleashing} & -- & -- & -- & 2.70 & DreamRec \cite{yang2024generate} & +21.62\% \\
\midrule
\multirow{4}{*}{KuaiRec \cite{gao2022kuairec}} & ADIGRec \cite{zhu2025adaptive} & 8.59 & DimeRec \cite{li2024dimerec} & +18.32\% & 5.31 & DreamRec \cite{yang2024generate} & +20.41\% \\
& TDM \cite{mao2025addressing} & 5.48 & DreamRec \cite{yang2024generate} & +5.48\% & 4.77 & DreamRec \cite{yang2024generate} & +13.84\% \\
 & TA-Rec \cite{mao2025efficiency} & 5.40 & DreamRec \cite{yang2024generate} & +4.65\% & 4.42 & DreamRec \cite{yang2024generate} & +4.42\% \\
 & DreamRec \cite{yang2024generate} & 5.16 & CL4SRec \cite{xie2022contrastive} & +21.41\% & 4.11 & CL4SRec \cite{xie2022contrastive} & +104\% \\
\midrule
 \multirow{5}{*}{Zhihu \cite{hao2021large}} & ADIGRec \cite{zhu2025adaptive} & 3.00 & DimeRec \cite{li2024dimerec} & +42.18\% & 0.99 & DimeRec \cite{li2024dimerec} & +30.26\% \\
 & TDM \cite{mao2025addressing} & 2.65 & DreamRec \cite{yang2024generate} & +14.72\% & 0.88 & DreamRec \cite{yang2024generate} & +10.23\% \\
 & TA-Rec \cite{mao2025efficiency} & 2.43 & DreamRec \cite{yang2024generate} & +7.52\% & 0.91 & PreferDiff \cite{liu2024preference} & +9.64\% \\
 & DreamRec \cite{yang2024generate} & 2.26 & CL4SRec \cite{xie2022contrastive} & +11.33\% & 0.79 & CL4SRec \cite{xie2022contrastive} & +6.76\% \\
 & DiffDiv \cite{cai2025unleashing} & -- & -- & -- & 0.86 & DreamRec \cite{yang2024generate} & +16.2\% \\

\bottomrule
\end{tabular}
\end{table}

\subsubsection{\textbf{Sequence as Diffusion Target and Guidance}} 
In sequential recommendation, the user interaction sequence can serve as both the target of the diffusion process and guidance in the reverse process. Diffusion can be applied to the entire sequence or to target item embeddings, with historical interactions providing contextual guidance \cite{huang2024dual, luo2025enhancing, chen2024semanticaware}. These methods are simple and effective for capturing local sequential patterns but may be limited in modeling broader user intent.

Other approaches incorporate user intent or interest representations by diffusing sequence-level embeddings and leveraging reference samples or aggregated interest vectors for reconstruction and contrastive learning \cite{qu2025intent, le2025diffusion}. Compared to direct sequence diffusion, these methods better capture global interests and user-level semantics, though they may require more sophisticated representation design.

The diffusion process can also operate on subsets of sequence items, using uncorrupted items as guidance to reconstruct corrupted ones, enabling bidirectional recovery, data augmentation, and positive view generation for contrastive learning \cite{liu2023diffusion, cui2024context, chen2025unlocking}. This enables bidirectional recovery, data augmentation, and positive view generation for contrastive learning, trading off some simplicity for richer learning signals.

Beyond representation learning, diffusion has been integrated into reinforcement learning frameworks to generate recommended items as actions \cite{chen2025energy}. By combining sequence corruption and guided reconstruction, these methods capture both local and global sequential patterns, improving the quality and diversity of recommendations.

\subsubsection{\textbf{Sequence as Diffusion Target}} 
Diffusion models can operate directly on user sequences for data augmentation, denoising, or representation learning. They can reconstruct the original sequence to generate augmented samples \cite{wu2023diff4rec}, denoise historical sequences deterministically or stochastically \cite{li2024multi, zheng2025feature, lee2024ediffurec}, compute noise levels from similar sequences to preserve semantic information, or generate sequence embeddings guided by context and entity information in conversational recommendation \cite{feng2025fusion, choi2025similarity, wang2026diffsbr, liu2025multi}. These approaches differ in complexity and modeling focus: direct sequence denoising is simple and effective for local patterns, while context-guided embedding diffusion captures richer semantic and temporal information but requires additional representation design.

Some methods function as plug-in modules in existing recommendation frameworks, generating soft positive augmentations, diffusing in alternative embedding spaces such as sememes, or pre-training sequence denoisers for downstream models \cite{ma2024plug, wang2023diffusion, ge2025time, ma2024seedrec, wu2024diffusion}. These provide flexibility and modularity, trading off implementation simplicity for enhanced augmentation or pre-training benefits.

Sequences can also serve as the final state of the diffusion process, where target items are diffused into representations of historical sequences using transformers or Brownian bridges, optionally conditioned on user embeddings or clustering information \cite{xie2024bridging, he2020lightgcn, bai2025unconditional}. Discrete diffusion has also been applied, with operations such as swapping or substituting items, or transitioning among semantic categories for reranking or next-item prediction tasks \cite{lin2024discrete, xie2024breaking, hou2023learning}. Compared to continuous diffusion in embedding space, discrete methods preserve the item's categorical structure, while continuous methods capture richer latent relationships.

Diffusion has additionally been applied for sequential CTR prediction. Sequence representations or interest vectors are reconstructed or diffused under guidance from queries, items, or related sequences to predict CTR and conversion \cite{li2025diffusiongs, lai2025modeling, zhao2025adaptive}. These approaches enable sequential recommendation models to capture temporal dependencies, semantic similarity, and user preferences more effectively.

\subsubsection{\textbf{Sequence as Diffusion Guidance}} In sequential recommendation, the historical user sequence can serve as guidance in the reverse diffusion process while the target item embedding is corrupted. Guidance is typically extracted from the sequence using neural architectures such as convolutional neural networks \cite{niu2024diffusion}, long short-term memory \cite{niu2025implicit}, or transformers \cite{wang2024leadrec, du2023sequential, li2023diffurec, yang2024generate}, with enhancements including cross-attention \cite{zolghadr2024generative}, user intention \cite{hu2024generate}, dynamic attention weights to capture evolving preferences \cite{liu2025generate}, or diversity-aware distributions \cite{cai2025unleashing, mao2025efficiency, zhu2025adaptive}. These approaches allow the reverse process to generate target item embeddings efficiently and accurately, often with DDIM sampling or one-step generation, but may be limited in capturing complex multi-interest patterns.

Other methods focus on enriching the guidance signal by capturing multi-interest and temporal user preferences \cite{li2024dimerec, wu2025learning}, leveraging semantic or categorical information \cite{mao2025distinguished}, improving robustness to missing or noisy data \cite{mao2025addressing}, and incorporating contrastive or preference-based objectives for stronger personalization \cite{song2025enhancing, wang2024conditional, liu2024preference, rafailov2024direct, liu2025alleviating}. Fairness-aware adjustments are also explored to improve equity in sequential recommendations \cite{liu2025alleviating}. These methods provide stronger personalization and robustness, but often require additional modeling complexity.

Diffusion guidance can additionally augment existing recommendation frameworks. For instance, it can capture user behavior drift to refine target item embeddings or generate auxiliary signals for embedding alignment in two-tower models \cite{wang2025unleashing}. By leveraging the historical sequence as structured guidance, these models improve the quality of generated embeddings, enabling more accurate and robust sequential recommendations.

\begin{figure*}[t]
\centering
\begin{tikzpicture}[
    root/.style={draw, rounded corners, minimum width=12mm, minimum height=8mm, fill=green!40, font=\bfseries\small, align=center},
    mid/.style={draw, rounded corners, minimum width=15mm, minimum height=3mm, font=\bfseries\small, align=center},
    cf/.style={mid, fill=red!40},
    sr/.style={mid, fill=blue!40},
    sub1/.style={draw, rounded corners, minimum width=70mm, minimum height=3mm, font=\footnotesize, align=left, text width=130mm, fill=#1},
    short/.style={draw, rounded corners, minimum width=70mm, minimum height=3mm, font=\footnotesize, align=left, text width=110mm, fill=#1},
    sub/.style={draw, rounded corners, minimum width=70mm, minimum height=8mm, font=\footnotesize, align=left, text width=105mm, fill=#1},
    arrow/.style={
    -{Latex},
    line width=0.7pt,
    draw=black!70
}
]

\node[root, rotate=90] (core) at (-3,1.7) {Data Modality and Domain};

\node[sub1=green!20] (image) at (5,3) {\textbf{Image Generation}: AdBooster \cite{shilova2023adbooster} CG4CTR \cite{yang2024new} Czapp et al. \cite{czapp2024dynamic} DiFashion \cite{xu2024diffusion} GEMRec \cite{guo2024gemrec} REBECA \cite{patron2025recommendations} CAIG \cite{chen2025ctr} VirtualModel \cite{chen2024virtualmodel} P\&R \cite{li2023planning} PerFusion \cite{lin2025sell} Vashishtha et al. \cite{vashishtha2024chaining} Pigeon \cite{xu2025personalized} RAGAR \cite{ling2025ragar} MMCRec \cite{mukande2024mmcrec} PMG \cite{shen2024pmg} GeneRec \cite{wang2023generative} Ramsey et al. \cite{ramsey2025cross}};
\node[sub1=green!20] (texttorec) at (5,2.1) {\textbf{Text-to-recommendation}: DMSR \cite{tomasi2024diffusion}};
\node[sub1=green!20] (mm) at (5,1.2) {\textbf{Multimodal Recommendation}: DiffMM \cite{jiang2024diffmm} LD4MRec \cite{yu2023ld4mrec} MCDRec \cite{ma2024multimodal} RDRec \cite{he2024diffusion} DiffCL \cite{song2025diffcl} CCDRec \cite{yang2025curriculum} DiffKD \cite{ma2025diffkd} ITCoHD-MRec \cite{hao2025itcohd} MDSBR \cite{li2025mdsbr} GDNSM \cite{ma2025generating} MoDiCF \cite{li2025generating} GenRec \cite{he2025flip} LSGM4Rec \cite{song2025boosting} KDiffE \cite{mo2025knowledge} Diff-MSIN \cite{cui2025diffusion} DCAR-DM \cite{xiu2026dual} AsymDiffRec \cite{zhu2025asymmetric} DMMD4SR \cite{lu2025dmmd4sr} M$^3$BSR \cite{cui2025multi}};
\node[sub1=green!20] (cd) at (5,0.2) {\textbf{Cross-domain Recommendation}: DiffCDR \cite{xuan2024diffusion} Diff-MSR \cite{wang2024diff} DMCDR \cite{li2025exploring} LLDCDR \cite{liu2025llm} HorizonRec \cite{zha2025align} DACDR \cite{li2025diffusion} CD-CDR \cite{li2025cd} FairCDR \cite{wu2025faircdr}};

\draw[arrow] (core.south) |- (image.west);
\draw[arrow] (core.south) |- (texttorec.west);
\draw[arrow] (core.south) |- (mm.west);
\draw[arrow] (core.south) |- (cd.west);

\end{tikzpicture}
\caption{Taxonomy on Data Modality and Domain.}
\label{fig:taxonomydata}
\end{figure*}

\begin{figure}
     \centering
     \begin{subfigure}{0.43\textwidth}
         \centering
         \includegraphics[scale=0.3]{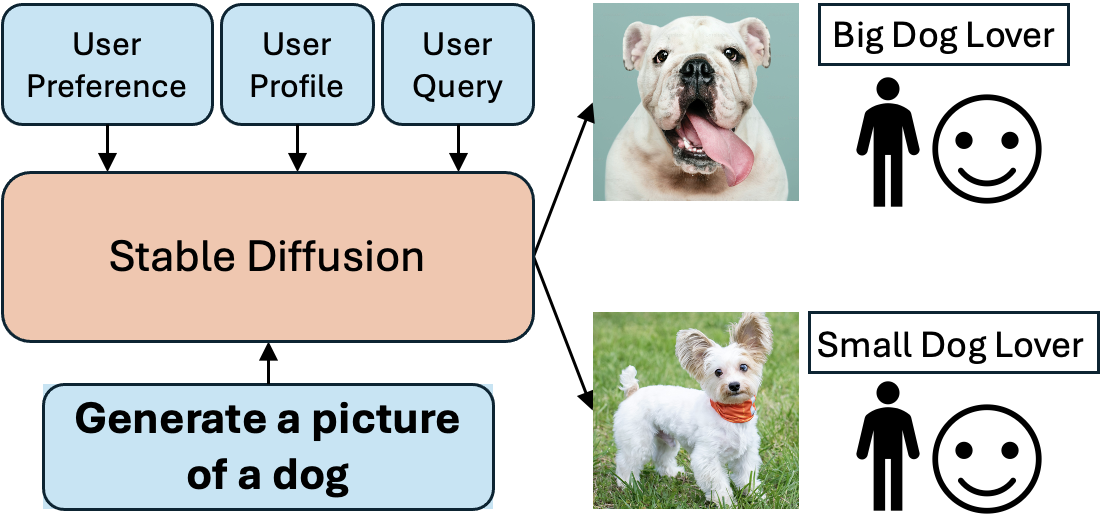}
         \caption{Image Generation}
     \end{subfigure}
     \hfill
     \begin{subfigure}{0.28\textwidth}
         \centering
         \includegraphics[scale=0.35]{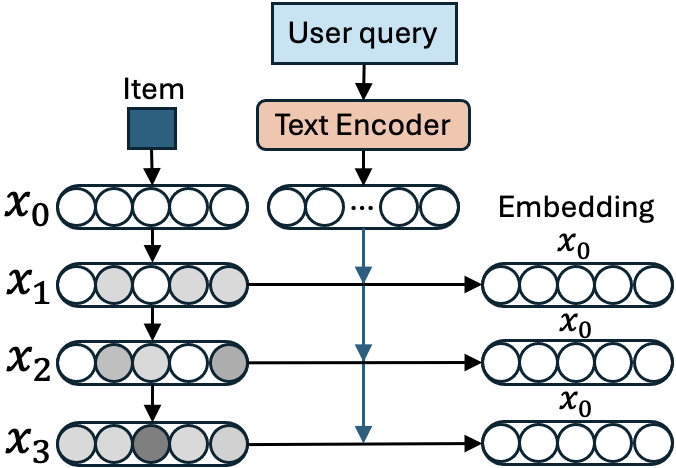}
         \caption{Text-to-rec Training}
     \end{subfigure}
     \hfill
     \begin{subfigure}{0.28\textwidth}
         \centering
         \includegraphics[scale=0.35]{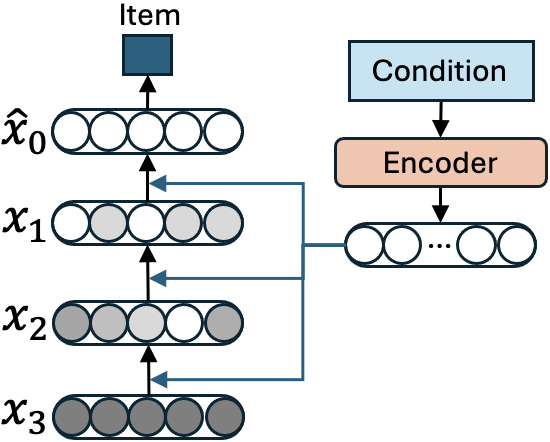}
         \caption{General Inference Framework}
     \end{subfigure}
        \caption{Multi-domain Recommendation with Diffusion Models. The pre-trained Stable Diffusion can generate product images tailored to the user's preference, profile and query. In text-to-rec, the user query is encoded to guide the denoising process of the target item embedding, which can be generalized to a framework that generates the target item embedding according to any condition.}
        \Description{Side by side plots.}
        \label{fig:mutlimodal}
\end{figure}


\section{Data Modality and Domain}
\label{sec:data}

Besides core recommendation tasks, another important axis concerns data modality and domain, where diffusion models operate across heterogeneous inputs and multiple domains. As illustrated in Fig.~\ref{fig:taxonomydata}, we categorize the related work into four directions: image generation, text-to-rec, multimodal recommendation, and cross-domain recommendation.

Fig.~\ref{fig:mutlimodal} provides an overview of how diffusion models interact with different modalities. For example, pre-trained Stable Diffusion models can generate product images tailored to a user’s preferences, profile, or query. In text-to-rec settings, user queries are encoded as conditions to guide the denoising of the target item embedding. More generally, this paradigm can be extended to a unified conditional generation framework, where the target item embedding is produced according to arbitrary side information or modality-specific signals.

\subsection{\textbf{Image Generation}} 
Diffusion models are integrated into recommender systems through image generation to enhance visualization, personalization, and advertising effectiveness. For personalized visualization, diffusion models generate outfit or product images in latent space by conditioning the reverse process on existing items and user history \cite{xu2024diffusion}. The same framework supports text-to-image generation for individual pieces, and feedback can further fine-tune generation with DPO \cite{yu2025fashiondpo}. Personalized image synthesis can also be guided by users’ historical ratings \cite{patron2025recommendations} or learned personalization tokens \cite{xu2025personalized}, while latent diffusion has been applied to cross-cultural settings \cite{ramsey2025cross}. Early personalization methods are simpler and effective for individual users, while feedback- or token-guided approaches capture richer user preferences but require more modeling and computational resources.

In advertising and CTR-oriented scenarios, diffusion models generate candidate product images under prompts conditioned on user group information \cite{yang2024new}, and predicted CTR are used to filter high-quality samples for fine-tuning Stable Diffusion \cite{rombach2022high} through low-rank adaptation \cite{hu2022lora}. Stable Diffusion is also fine-tuned directly for product promotion \cite{shilova2023adbooster}. Structural constraints such as object edges can be incorporated via ControlNet \cite{zhang2023adding} to generate optimized backgrounds before reinserting the original object \cite{czapp2024dynamic}. Reward models trained to estimate CTR or user preference further guide background or product image generation \cite{chen2025ctr, lin2025sell}. Similar techniques are applied to automated banner and marketing content creation \cite{vashishtha2024chaining}. Compared to personalization-focused generation, these methods emphasize advertising effectiveness and structural fidelity, trading off some flexibility for task-specific optimization.

Diffusion-based generation can also be embedded into the recommendation loop. Generated images allow interactive ranking based on user engagement \cite{guo2024gemrec}. Retrieval-augmented diffusion supports personalized micro-video thumbnail generation conditioned on historical content \cite{wang2023generative, blattmann2022retrieval}. Optimization through pretrained multimodal ranking models \cite{zhang2022latent} with policy gradients \cite{sutton1999policy} further aligns generated images with user preferences \cite{ling2025ragar}. In conversational systems, diffusion models generate multimodal responses including images, audio, and video \cite{mukande2024mmcrec}, and generated images can also supervise embedding learning \cite{shen2024pmg}. Controllable content creation methods, e.g., layout-aware product placement \cite{li2023planning} and pose-constrained human-object interaction \cite{chen2024virtualmodel}, offer fine-grained structure at the cost of more complex modeling.

\subsection{\textbf{Text-to-Recommendation}} Instead of modeling a user’s historical interactions, DMSR \cite{tomasi2024diffusion} addresses slate recommendation by generating a list of items solely based on a user query. The diffusion model operates in the item embedding space, where an entire slate (e.g., a music playlist) is diffused and reconstructed in the reverse process conditioned on a textual description. At inference, a playlist is generated by iteratively applying the reverse diffusion process given the query. Unlike many prior approaches that predict noise or embeddings directly, this method adopts v-prediction \cite{salimans2022progressive}, leading to improved convergence and empirical performance. Generated embeddings are mapped to the nearest candidate items.

\subsection{\textbf{Multimodal Recommendation}} 
Multimodal recommendation incorporates additional attributes such as visual, textual, acoustic, or behavioral signals into the diffusion process. A common strategy is to denoise multimodal representations or interaction structures to obtain more reliable embeddings for recommendation \cite{jiang2024diffmm, ma2025diffkd, hao2025itcohd, xiu2026dual, mo2025knowledge, lu2025dmmd4sr, li2025mdsbr, cui2025multi}. The denoising process can be applied to user interaction sequences, user-item graphs, modality-specific embeddings, or knowledge graphs between items and their attributes, and in some cases is performed in multiple stages to address domain shift or multi-behavior settings \cite{lu2025dmmd4sr, cui2025multi}. Conditioning on other modalities can further enhance robustness \cite{cui2025diffusion}. Simple denoising methods are efficient and effective for single-modality signals, while cross-modality conditioning better captures complex interactions but increases modeling complexity.

Beyond representation denoising, diffusion is used to construct augmented views for contrastive learning \cite{song2025diffcl} or to complete missing modalities \cite{li2025generating}. External encoders such as LLMs and CLIP \cite{radford2021learning} can further enhance multimodal representations before being incorporated into the diffusion framework \cite{song2025boosting}. Compared to standard denoising, these methods capture richer multimodal semantics but require additional pretraining or feature extraction.

Instead of only refining representations, diffusion models can directly generate recommendation outputs in multimodal settings. The interaction sequence can be diffused and reconstructed into a recommendation list while conditioning on multimodal item features \cite{yu2023ld4mrec}. Item embeddings can also be diffused under visual and textual guidance and integrated into ranking objectives \cite{ma2024multimodal}. Iterative denoising further enables curriculum-style negative sampling \cite{yang2025curriculum} or modality-aware negative generation \cite{ma2025generating}. In rating prediction scenarios, user-item embeddings reconstructed from review-based features can be denoised from Gaussian noise and used for final score estimation \cite{he2024diffusion}. Direct output generation captures richer user-item interactions but generally involves higher computational and implementation overhead compared to representation-only methods.

In addition to continuous diffusion, discrete diffusion has also been explored in multimodal recommendation, where features or interaction structures are modified through discrete operations and reconstructed in latent space for CTR \cite{zhu2025asymmetric} or graph-based recommendation \cite{he2025flip}. Compared to continuous diffusion, discrete approaches better preserve categorical or structural information, while continuous methods offer richer latent modeling.

\subsection{\textbf{Cross-domain Recommendation}}
Diffusion models have also been applied to cross-domain recommendation, where discrepancies between domains may cause negative transfer and exacerbate the cold-start problem. One line of work leverages diffusion for data generation, producing informative samples to improve model adaptation in cold-start domains \cite{wang2024diff}. These methods are straightforward and effective for augmenting sparse domains but may increase training overhead.

Rather than generating additional training data, other approaches directly synthesize user representations for the target domain by conditioning on preferences learned in the source domain, often using classifier-free guidance to control generation \cite{xuan2024diffusion, li2025exploring}. Compared to data-generation methods, representation synthesis directly transfers knowledge without expanding the dataset, trading off simplicity for more controlled adaptation. Stability and efficiency are further improved through auxiliary rating prediction and accelerated solvers \cite{xuan2024diffusion, lu2022dpm}. 

Beyond representation synthesis, diffusion also facilitates cross-domain transfer through knowledge enhancement and latent alignment, including incorporating LLM knowledge \cite{liu2025llm}, diffusing user representations under adaptive schedules in sequential settings \cite{zha2025align}, aligning shared user embeddings across domains \cite{li2025diffusion}, and generating target-domain item embeddings from unified user representations with domain indicators \cite{li2025cd}. These approaches offer richer cross-domain modeling and alignment at the cost of increased complexity. Overall, diffusion models provide flexible mechanisms for generation, alignment, and representation transfer, helping mitigate negative transfer and improving recommendation performance across domains.

\section{Trustworthy Objectives and Constraints}
\label{sec:trustworthy}
Beyond optimizing recommendation performance, diffusion models have increasingly been explored under the lens of trustworthy objectives and constraints, addressing key concerns such as fairness, accountability, transparency, and OOD robustness. As illustrated in Fig.~\ref{fig:taxonomytrustworthy}, we organize the related literature into four main directions corresponding to these trustworthy aspects.

\subsection{\textbf{Fairness}} 
Diffusion models have been applied to enhance fairness in recommender systems, addressing biases that can discriminate against specific user groups. By incorporating user group information into the reverse diffusion process, rating predictions can be adjusted to mitigate unfairness \cite{jiang2024diffairec}, and analyses show that certain diffusion architectures may require modification to improve fairness \cite{malitesta2024fair}. These approaches are simple to integrate but may be limited in addressing more complex biases.

Beyond group fairness, diffusion-based approaches also tackle popularity bias and improve diversity. Synthetic sessions can be generated to reduce the over-recommendation of popular items \cite{gupta2024guided, gupta2024scm4sr}, while user preferences and embeddings can be reweighted to better reflect underrepresented items \cite{he2024balancing}. Category-level guidance in the diffusion process allows control over the diversity of recommended items, enabling the selection of items from specific categories while iteratively denoising the interaction vector \cite{han2024controlling}. Compared to basic group-based adjustments, these techniques better capture multi-faceted fairness objectives but often involve additional modeling complexity.

These fairness techniques extend to cross-domain and sequential recommendation scenarios. Diffusion models can align embeddings across domains \cite{wu2025faircdr}, enforce fairness constraints over sequences \cite{liu2025alleviating}, or incorporate adversarial regularization to ensure that both user and item representations are fair and balanced \cite{yang2025adversarial}, promoting more equitable recommendations. These methods provide stronger equity guarantees and robustness across recommendation contexts, trading off simplicity for comprehensive fairness control.

\begin{figure*}[t]
\centering
\begin{tikzpicture}[
    root/.style={draw, rounded corners, minimum width=12mm, minimum height=8mm, fill=cyan!40, font=\bfseries\scriptsize, align=center},
    mid/.style={draw, rounded corners, minimum width=15mm, minimum height=3mm, font=\bfseries\small, align=center},
    cf/.style={mid, fill=red!40},
    sr/.style={mid, fill=blue!40},
    sub1/.style={draw, rounded corners, minimum width=70mm, minimum height=3mm, font=\footnotesize, align=left, text width=130mm, fill=#1},
    short/.style={draw, rounded corners, minimum width=70mm, minimum height=3mm, font=\footnotesize, align=left, text width=110mm, fill=#1},
    sub/.style={draw, rounded corners, minimum width=70mm, minimum height=8mm, font=\footnotesize, align=left, text width=105mm, fill=#1},
    arrow/.style={
    -{Latex},
    line width=0.7pt,
    draw=black!70
}
]

\node[root, rotate=90] (core) at (-3,1.5) {Trustworthy Objectives and Constraints};

\node[sub1=cyan!20] (fairness) at (5,3) {\textbf{Fairness}: CGSoRec \cite{he2024balancing} D3Rec \cite{han2024controlling} DCASR \cite{gupta2024guided} DifFaiRec \cite{jiang2024diffairec} Malitesta et al. \cite{malitesta2024fair} FairCDR \cite{wu2025faircdr} DiffuFair \cite{yang2025adversarial} FairGENRec \cite{liu2025alleviating}};
\node[sub1=cyan!20] (accountability) at (5,1.9) {\textbf{Accountability}: GDMPD \cite{yuan2023manipulating}  IPDGI \cite{chen2023adversarial} PaDiRec \cite{shen2024generating} SDRM \cite{lilienthal2024multi} ToDA \cite{liu2024toda} DGFedRS \cite{di2024federated} LDM \cite{chen2025latent} FedDiffRec \cite{li2025feddiffrec} Diff-WassGAN \cite{le2025adversarial} PRISM \cite{ma2025bridging} DLDA \cite{qiao2025controllable} DPDM \cite{wang2025differentially}};
\node[sub1=cyan!20] (transparency) at (5,0.9) {\textbf{Transparency}: DIEXRS \cite{guo2023explainable}};
\node[sub1=cyan!20] (ood) at (5,0) {\textbf{Out-of-distribution}: CausalDiffRec \cite{zhao2024graph} DRGO \cite{zhao2025distributionally} DiffDRG \cite{bao2025distributed} CNSDiff \cite{zhao2025causal}};

\draw[arrow] (core.south) |- (fairness.west);
\draw[arrow] (core.south) |- (accountability.west);
\draw[arrow] (core.south) |- (transparency.west);
\draw[arrow] (core.south) |- (ood.west);

\end{tikzpicture}
\caption{Taxonomy on Trustworthy Objectives and Constraints.}
\label{fig:taxonomytrustworthy}
\end{figure*}

\subsection{\textbf{Accountability}} 
Beyond optimizing recommendation accuracy, diffusion models have been leveraged to enhance privacy, security, and robustness in recommender systems. They can generate synthetic user interactions to replace sensitive data, improving privacy and resisting malicious attacks \cite{lilienthal2024multi, liu2024toda, chen2025latent}. By diffusing user profiles or product images in latent space, models can produce fake data for testing or defend against shilling attacks, while reverse processes condition on target items or image embeddings to preserve realism \cite{qiao2025controllable, le2025adversarial, chen2023adversarial}. These methods are straightforward and effective for privacy and attack resistance, though they require careful design to maintain realism.

Diffusion models also enable flexible and controllable recommendation objectives. They can generate model parameters or synthetic data aligned with specific goals such as accuracy, diversity, or fairness \cite{shen2024generating}, or incorporate signals like news authenticity \cite{ma2025bridging}. Additionally, diffusion can perturb item rating matrices to ensure differential privacy during data transfer \cite{wang2025differentially}. Compared to basic synthetic data generation, these methods provide more precise control over recommendation outcomes but often involve additional complexity.

Diffusion models have also been adapted to federated recommender systems \cite{li2025feddiffrec}, enabling local reconstruction of user sequences while maintaining global model privacy. Iterative forward and reverse processes generate augmented sequences for sequential recommendation \cite{di2024federated}. Separately, diffusion models have been applied to defend against attacks on product data. Pre-trained diffusion models can purify product images to detect promotion attacks by denoising and comparing embeddings with the original images \cite{yuan2023manipulating}. These approaches trade off modeling and computational complexity for stronger privacy, security, and robustness guarantees.

\subsection{\textbf{Transparency}} 
Transparency is a key component of responsible recommender systems. Diffusion models can support this by generating interpretable explanations alongside recommendations. For instance, DIEXRS \cite{guo2023explainable} leverages user, item, and product review embeddings to produce recommendation explanations. A transformer model predicts and denoises the noise added to these embeddings, while a decoder generates human-readable explanations. Through iterative denoising, the model recovers sentence embeddings that are then used by the decoder to explain why a particular item is recommended.


\subsection{\textbf{Out-of-Distribution}} 
OOD recommendation tackles shifts in user preferences or item interactions due to temporal, exposure, or popularity changes, which can introduce bias and reduce fairness. Standard recommenders often fail to generalize under such distributional shifts. Diffusion models help improve robustness by denoising training data or embeddings. Some methods remove noise from user interactions to enhance distributionally robust optimization \cite{zhao2025distributionally}, which is simple and effective but may focus only on local noise patterns.

Graph-based approaches generate multiple versions of user-item graphs and reconstruct embeddings conditioned on latent factors, facilitating invariant representations for downstream models \cite{zhao2024graph, he2020lightgcn}. Other techniques focus directly on graph embeddings, applying stepwise denoising to improve generalization \cite{bao2025distributed} or generating negative samples of varying difficulty to strengthen OOD performance \cite{zhao2025causal}. Compared to interaction-level denoising, these graph-based methods capture richer relational and structural information but often require higher computational cost. These strategies demonstrate how diffusion models can purify data and enrich embeddings, enabling recommendation systems to better handle shifting distributions and evolving user behavior.

\section{Datasets}
\label{sec:datasets}
Recommender systems with diffusion models have been studied in various application domains. We compile the datasets used and categorize them based on their domains, including:

\begin{itemize}
    \item \textbf{Fashion and E-commerce} datasets that focus on fashion and product recommendations for shopping platforms.

    \item \textbf{Social and Review-based} datasets that express user preferences through ratings, reviews, and social connections.

    \item \textbf{Media and Entertainment} datasets that cover music, movies, and other media consumption patterns.

    \item \textbf{Location-based} datasets that capture users’ check-in behavior and spatial-temporal preferences.

    \item \textbf{Meal Planning} datasets that focus on recipe recommendations and user feedback.
\end{itemize}

\subsection{Fashion and E-commerce Datasets}

The Fashion and E-commerce domain includes datasets capturing rich multimodal information such as product images, textual metadata, and user interaction logs. Representative datasets include Ali-CCP \cite{ma2018entire}, Alibaba-iFashion \cite{chen2019pog}, Amazon \cite{mcauley2013hidden, he2016ups, mcauley2015image, ni2019justifying, hou2024bridging}, Ciao \cite{tang2012mtrust}, CreativeRanking \cite{wang2021hybrid}, HoIHuman \cite{chen2024virtualmodel}, M5Product \cite{dong2022m5product}, Polyvore \cite{lu2019tk}, PPG30K \cite{li2023planning}, RetailRocket \cite{yang2022multi}, Tmall \cite{zhang2014large, zhong2015stock}, and YooChoose \cite{ben2015recsys}. Table \ref{tab:fashion_datasets} summarizes their main statistics and characteristics.

These datasets vary in scale, modality, and application focus. Some emphasize product interactions and purchase histories (e.g., Ali-CCP, Amazon, Tmall, YooChoose), while others provide rich visual or multimodal content such as fashion images, human-object interactions, and textual metadata (e.g., Alibaba-iFashion, HoIHuman, M5Product, Polyvore). Certain datasets additionally incorporate advertising impressions, outfit compositions, or social review elements (e.g., CreativeRanking, Ciao, PPG30K). Together, they provide diverse settings for evaluating recommendation models on tasks including purchase prediction, outfit composition, and visual-textual understanding.

\begin{table*}
\small
\centering
  \caption{Fashion and E-commerce Datasets}
  \label{tab:fashion_datasets}
  \rowcolors{2}{gray!15}{white}
  \begin{tabular}{>{\raggedright\arraybackslash}m{3cm}rr>{\centering\arraybackslash}m{7cm}}
    \toprule
    \textbf{Dataset} & \textbf{Users} & \textbf{Size} & \textbf{Used in}\\
    \midrule
    Ali-CCP \cite{ma2018entire} & 400K & 3,400,000 interactions & \cite{wu2025faircdr} \\
    Alibaba-iFashion \cite{chen2019pog} & 3.57M & 1,781,093 interactions & \cite{jiang2024diffkg, xu2024diffusion, li2025divide, li2025disentangled, bui2025personalized, cui2025rsdiff, yu2025fashiondpo, ling2025ragar, shen2024pmg, zhang2025modeling, chen2025dual}\\
    Amazon \cite{mcauley2013hidden, he2016ups, mcauley2015image, ni2019justifying, hou2024bridging} & 54M & 571,540,000 reviews & \cite{chen2023adversarial, guo2023explainable, wang2024conditional, zhao2024denoising, dong2024dices, wang2024diff, wu2023diff4rec, jiang2024diffmm, li2023diffurec, liu2023diffusion, xuan2024diffusion, he2024diffusion, cui2024context, lee2024ediffurec, zhu2024graph, yu2023ld4mrec, lilienthal2024multi, ma2024multimodal, ma2024plug, walker2022recommendation, du2023sequential, wang2023diffusion, ma2024seedrec, wang2024leadrec, shen2024generating, wu2024diffusion, di2024federated, xie2024bridging, lee2024stochastic, choi2023blurring, buchanan2024incorporating, huang2024dual, liu2024preference, chen2024g, xie2024breaking, chen2024semanticaware, hu2024generate, yuan2023manipulating, lee2025collaborative, li2025exploring, song2025diffcl, zolghadr2024generative, yuan2025hyperbolic, chen2025conditional, qu2025generative, you2024context, hao2025diff, le2025diffusion, wu2025learning, chen2025unlocking, zhang2025graph, li2024diffgca, luo2025enhancing, yang2025curriculum, mao2025distinguished, mao2025addressing, ma2025diffkd, qu2025intent, zheng2025feature, le2025adversarial, hao2025itcohd, li2025mdsbr, ma2025generating, liu2025llm, zha2025align, li2025generating, liu2025generate, he2025flip, song2025boosting, li2025diffusion, mo2025knowledge, liu2025alleviating, jing2025interest, choi2025similarity, shi2025dagr, cui2025diffusion, lu2025dmmd4sr, xiu2026dual, feng2025fusion, li2025cd, ni2025hierdiffuse, hu2025fading}\\
    Ciao \cite{tang2012mtrust} & 12K & 484,086 ratings & \cite{li2024recdiff, he2024balancing, zang2025diffusion, liu2024score, li2024graph, sun2025model, li2025feature, gao2025graph}\\
    CreativeRanking \cite{wang2021hybrid} & - & 1,204,988 images & \cite{yang2024new, chen2025ctr} \\
    HoIHuman \cite{chen2024virtualmodel} & - & 3,000,000 images & \cite{chen2024virtualmodel}\\
    M5Product \cite{dong2022m5product} & - & 6,313,067 samples & \cite{mukande2024mmcrec}\\
    Polyvore \cite{lu2019tk} & 630 & 127,326 outfits & \cite{xu2024diffusion, yu2025fashiondpo}\\
    PPG30K \cite{li2023planning} & - & 34,150 images & \cite{li2023planning}\\
    RetailRocket \cite{yang2022multi} & 11K & 87,822 interactions & \cite{li2025diffgraph, song2025enhancing, mo2026hierarchical, wang2026diffsbr} \\
    Tmall \cite{zhang2014large, zhong2015stock} & - & 1,333,729,303 interactions & \cite{li2024multi, lee2024stochastic, hao2025diff, li2025diffgraph, zheng2025diffusion, cui2025diffusion, cui2025multi, mo2026hierarchical, chen2025causal}\\
    YooChoose \cite{ben2015recsys} & - & 11,825,218 sessions & \cite{niu2024diffusion, yang2024generate, li2024dimerec, chen2024semanticaware, niu2025implicit, mao2025addressing, zhu2025adaptive, mao2025efficiency, bai2025unconditional, cai2025unleashing}\\
  \bottomrule
\end{tabular}
\end{table*}

\subsection{Social and Review-based Datasets}

The Social and Review-based domain includes platforms where user preferences are expressed through explicit feedback such as ratings and reviews, often accompanied by social interactions or user-generated content. Representative datasets include Dianping \cite{li2015overlapping}, Douban \cite{zhu2019dtcdr}, Epinions \cite{fan2019graph}, Yelp, and Zhihu \cite{hao2021large}. Table~\ref{tab:social_datasets} summarizes their main statistics and characteristics.

These datasets typically combine user-item interactions with additional relational or textual information. Some platforms provide explicit social or trust connections (e.g., Dianping and Epinions), while others span multiple content domains (e.g., Douban) or include large-scale review and question-answer data (e.g., Yelp and Zhihu). Together, they offer diverse settings for evaluating recommendation models in socially enriched scenarios.

\begin{table*}
\small
\centering
  \caption{Social and Review-based Datasets}
  \label{tab:social_datasets}
  \rowcolors{2}{gray!15}{white}
  \begin{tabular}{>{\raggedright\arraybackslash}m{3cm}rr>{\centering\arraybackslash}m{7cm}}
    \toprule
    \textbf{Dataset} & \textbf{Users} & \textbf{Size} & \textbf{Used in}\\
    \midrule
    Dianping \cite{li2015overlapping} & 147K & 2,149,675 ratings & \cite{liu2024score} \\
    Douban \cite{zhu2019dtcdr} & - & 1,444,151 interactions & \cite{wang2024diff, zhao2024graph, ma2024plug, lee2024stochastic, he2024balancing, li2024graph, li2025dual, sun2025model, qiao2025controllable, zha2025align, zhang2025gdiffmae, gao2025graph, bao2025distributed, li2025cd}\\
    Epinions \cite{fan2019graph} & 18K & 764,352 ratings & \cite{li2024recdiff, zang2025diffusion, liu2024score, sun2025model}\\
    Yelp & 1M & 6,990,280 reviews & \cite{yi2024directional, guo2023explainable, hou2024collaborative, zhao2024denoising, dong2024dices, wu2023diff4rec, liu2023diffusion, cui2024context, zhao2024graph, zhu2024graph, li2024recdiff, wang2023diffusion, di2024federated, xie2024bridging, lee2024stochastic, choi2023blurring, buchanan2024incorporating, zang2025diffusion, huang2024dual, lee2025collaborative, yuan2025hyperbolic, wu2025learning, chen2025unlocking, peng2024diffusion, ju2025diffgr, zhang2025graph, yang2024diffgcl, li2025dual, liu2025ggdhscl, zhao2025distributionally, le2025adversarial, li2025mdsbr, zhao2025causal, liu2025alleviating, jing2025interest, choi2025similarity, zhang2025gdiffmae, li2025feature, feng2025fusion, wen2026condiff, ge2025time, bao2025distributed, li2025mask}\\
    Zhihu \cite{hao2021large} & 798K & 99,978,523 interactions & \cite{yang2024generate, chen2024semanticaware, niu2025implicit, mao2025addressing, zhu2025adaptive, mao2025efficiency, cai2025unleashing}\\
    
  \bottomrule
\end{tabular}
\end{table*}

\begin{table*}
\small
\centering
  \caption{Media and Entertainment Datasets}
  \label{tab:media_datasets}
  \rowcolors{2}{gray!15}{white}
  \begin{tabular}{>{\raggedright\arraybackslash}m{3cm}@{} r r  r >{\centering\arraybackslash}m{6cm}}
    \toprule
    \textbf{Dataset} & \textbf{Domain} & \textbf{Users} & \textbf{Size} & \textbf{Used in} \\
    \midrule
    FilmTrust \cite{guo2013novel} & Movies & 1.5K & 35,497 ratings & \cite{liu2024toda} \\
    MovieLens-100K \cite{harper2015movielens} & Movies & 1K & 100,000 ratings & \cite{jiangzhou2024dgrm, lilienthal2024multi, priyam2024edge, chen2025conditional, chen2025unlocking, peng2024diffusion, le2025adversarial, shen2024pmg, qiao2025controllable, wang2025differentially}\\
    MovieLens-1M \cite{harper2015movielens} & Movies& 6K & 1,000,209 ratings & \cite{yi2024directional, hou2024collaborative, zhao2024denoising, jiangzhou2024dgrm, dong2024dices, li2023diffurec, cui2024context, zhu2024graph, lilienthal2024multi, benedict2023recfusion, walker2022recommendation, du2023sequential, wang2023diffusion, wang2024leadrec, wu2023diff4rec, shen2024generating, di2024federated, lee2024stochastic, buchanan2024incorporating, jiang2024diffairec, priyam2024edge, malitesta2024fair, chen2024g, chen2024semanticaware, hu2024generate, han2024controlling, yuan2023manipulating, zolghadr2024generative, yuan2025hyperbolic, chen2025conditional, you2024context, wu2025learning, peng2024diffusion, ju2025diffgr, zhang2025graph, wang2025unleashing, luo2025enhancing, li2025feddiffrec, mao2025distinguished, qu2025intent, zheng2025feature, zhu2025addressing, le2025adversarial, qiao2025controllable, liu2025generate, liu2025alleviating, wang2025differentially, choi2025similarity, chen2025energy, zhang2025gdiffmae, wen2026condiff, yang2025adversarial, ge2025time, li2025mask, hu2025fading}\\
    MovieLens-10M \cite{harper2015movielens} & Movies& 72K & 10,000,054 ratings & \cite{li2024dimerec}\\
    MovieLens-20M \cite{harper2015movielens} & Movies& 138K & 20,000,263 ratings & \cite{walker2022recommendation}\\
    MovieLens-25M \cite{harper2015movielens} & Movies& 162K & 25,000,095 ratings & \cite{benedict2023recfusion, han2025diffusion, shi2025dagr}\\
    Netflix \cite{bennett2007netflix} & Movies& 480K & 100,000,000 ratings & \cite{benedict2023recfusion}\\
    \midrule
    Last-FM \cite{bertinmahieux2011} & Music& - & 1,000,000 songs & \cite{jiang2024diffkg, he2024balancing, jiang2024diffairec, chen2025conditional, qu2025generative, yang2024diffgcl, li2025feddiffrec, li2024graph, liu2025ggdhscl, cui2025rsdiff, yang2025adversarial, li2025mask} \\
    Netease \cite{cao2017embedding} & Music& 18K & 1,128,065 interactions & \cite{li2025divide, zhang2025modeling, chen2025dual} \\
    Spotify \cite{chen2018recsys}& Music & - & 1,000,000 playlists & \cite{tomasi2024diffusion} \\
    \midrule
    KuaiRand \cite{gao2022kuairand} & Visuals& 27K & 322,278,385 interactions & \cite{wang2025unleashing, chen2025energy} \\ 
    KuaiRec \cite{gao2022kuairec} & Visuals& 72K & 12,530,806 interactions & \cite{niu2024diffusion, yang2024generate, zhao2024graph, li2024dimerec, wang2024leadrec, niu2025implicit, wu2025learning, luo2025enhancing, mao2025addressing, zhao2025distributionally, zhu2025adaptive, zhao2025causal, chen2025energy, mao2025efficiency, bai2025unconditional}\\
    PickaPic \cite{kirstain2023pick} & Visuals& 4K & 583,747 pairs & \cite{lin2025sell} \\
    PixelRec \cite{cheng2024image} & Visuals & 29M & 195,755,320 interactions & \cite{ma2024seedrec}\\
    \midrule
    Steam \cite{kang2018self, wan2018item, pathak2017generating} & Games& 2M & 7,793,069 reviews & \cite{wu2023diff4rec, li2023diffurec, liu2023diffusion, hu2024generate, han2024controlling, li2025gddrec, mao2025distinguished, mao2025addressing, zheng2025feature, hu2025fading} \\
    \midrule
    Youshu \cite{chen2019matching} & Books & 8K & 51,377 interactions & \cite{li2025divide, bui2025personalized, zhang2025modeling, chen2025dual} \\
    \midrule
    MIND \cite{wu2020mind} & News & 1M & 24,155,470 interactions & \cite{dong2024dices, jiang2024diffkg, wu2024diffusion, cui2025rsdiff, li2025mask} \\

    
  \bottomrule
\end{tabular}
\end{table*}

\subsection{Media and Entertainment Datasets}
Datasets in the media and entertainment domain capture user interactions with movies, music, visuals, games, books, and news, reflecting preferences, consumption patterns, and content engagement behaviors. Table \ref{tab:media_datasets} summarizes their main statistics and characteristics.


\textbf{Movie Datasets} include FilmTrust \cite{guo2013novel}, MovieLens \cite{harper2015movielens}, and Netflix \cite{bennett2007netflix}. They provide explicit user ratings, often with additional metadata such as user demographics and movie genres. These datasets vary in scale from thousands of ratings to hundreds of millions, enabling research on collaborative filtering, rating prediction, and recommendation evaluation under different sparsity settings.

\textbf{Music datasets} include Last-FM \cite{bertinmahieux2011}, Netease \cite{cao2017embedding}, and Spotify \cite{chen2018recsys}. They cover user-song and user-playlist interactions, with some providing tag or similarity information, playlist completion tasks, and associated metadata. Together, they enable modeling of user preferences, sequential behaviors, and multi-level music recommendation challenges.





\textbf{Visual datasets} include KuaiRand \cite{gao2022kuairand}, KuaiRec \cite{gao2022kuairec}, PickaPic \cite{kirstain2023pick}, and PixelRec \cite{cheng2024image}. They capture user interactions with videos or images, often including multimodal information such as user comments, textual prompts, or cover images. These datasets support research in video and image recommendation, multimodal learning, and preference modeling for visual content.






\textbf{Game datasets} such as Steam \cite{kang2018self, wan2018item, pathak2017generating} collect user reviews, purchase histories, and bundle interactions. They provide both rating and behavioral signals, making them suitable for exploring recommendation tasks in gaming, bundle prediction, and user engagement modeling.



\textbf{Book datasets} like Youshu \cite{chen2019matching} contain user interactions with books and book bundles. They capture review behavior and bundle composition preferences, providing a basis for studying recommendation under item grouping and bundle-aware strategies.



\textbf{News datasets} such as MIND \cite{wu2020mind} include impression logs, clicks, and article metadata (title, abstract, body, and category). These datasets are designed for news recommendation tasks, supporting research on sequential behavior modeling, content-based recommendation, and evaluation of click prediction models.



\subsection{Location-based Datasets}

Location-based datasets capture users’ spatial-temporal behaviors through check-ins, providing insights into movement patterns, geographical preferences, and social mobility. Representative datasets include Foursquare \cite{yang2016participatory, yang2015nationtelescope}, Gowalla \cite{cho2011friendship}, and WeePlace \cite{liu2013personalized}. Table \ref{tab:pos_datasets} summarizes their main statistics and characteristics.

These datasets vary in scale, geographic coverage, and social connectivity. Foursquare covers millions of check-ins across hundreds of cities worldwide, Gowalla includes social friendship links along with check-in behavior, and WeePlace provides location categories and rich contextual information.





\begin{table*}
\small
\centering
  \caption{Location-based Datasets}
  \label{tab:pos_datasets}
  \rowcolors{2}{gray!15}{white}
  \begin{tabular}{>{\raggedright\arraybackslash}m{3cm}rr>{\centering\arraybackslash}m{7cm}}
    \toprule
    \textbf{Dataset} & \textbf{Users} & \textbf{Size} & \textbf{Used in}\\
    \midrule
    Foursquare \cite{yang2016participatory, yang2015nationtelescope} & 266K & 33,278,683 check-ins & \cite{qin2023diffusion, yi2024directional, long2024diffusion, wang2024dsdrec, zuo2024diff, malitesta2024fair, pan2025hgdrec, song2025enhancing, li2025dmsdrec, zuo2025bridging}\\
    Gowalla \cite{cho2011friendship} & 196K & 6,442,890 check-ins & \cite{qin2023diffusion, liu2024toda, lee2024stochastic, choi2023blurring, hao2025diff, le2025diffusion, peng2024diffusion, pan2025hgdrec, liu2025ggdhscl, song2025enhancing, zuo2025bridging, jing2025interest}\\
    WeePlace \cite{liu2013personalized} & 4K & 923,600 check-ins & \cite{long2024diffusion}\\
  \bottomrule
\end{tabular}
\end{table*}

\subsection{Meal Planning Datasets}

Meal planning datasets capture user interactions with recipes, meals, and ingredients, supporting research in recipe recommendation and meal composition modeling. Representative datasets include Allrecipes \cite{gao2019hierarchical}, Food \cite{majumder-etal-2019-generating}, and MealRec \cite{li2024mealrec}. Table \ref{tab:food_datasets} summarizes their main statistics and characteristics.

These datasets vary in granularity, from individual recipes and ingredients (e.g., Allrecipes, Food) to bundled courses and meal-level interactions (e.g., MealRec). They include textual, ingredient, and image information for evaluating recommendation models in the culinary domain.





\begin{table*}
\small
\centering
  \caption{Meal Planning Datasets}
  \label{tab:food_datasets}
  \rowcolors{2}{gray!15}{white}
  \begin{tabular}{>{\raggedright\arraybackslash}m{3cm}rr>{\centering\arraybackslash}m{7cm}}
    \toprule
    \textbf{Dataset} & \textbf{Users} & \textbf{Size} & \textbf{Used in}\\
    \midrule
    Allrecipes \cite{gao2019hierarchical} & 68K & 1,093,845 interactions & \cite{li2025generating} \\
    Food \cite{majumder-etal-2019-generating} & 25K & 718,379 reviews & \cite{zhao2024graph, zhao2025distributionally, zhao2025causal, bao2025distributed}\\
    MealRec \cite{li2024mealrec} & 2K & 181,087 interactions & \cite{li2025disentangled, bui2025personalized, zhang2025modeling} \\
  \bottomrule
\end{tabular}
\end{table*}

\section{Open Research Directions}
\label{sec:open}

\subsection{Efficiency}

Early diffusion models are computationally intensive, requiring hundreds or thousands of denoising steps to generate a single sample. In the generative modeling literature, multiple strategies have been proposed to alleviate this issue, including DDIM sampling \cite{song2020denoising}, latent-space diffusion \cite{rombach2022high}, and ODE-based solvers for score-based models \cite{song2020score}. 

Efficiency is particularly critical in recommender systems, where online inference latency directly affects user experience and system scalability \cite{hou2025generative}. Diffusion-based recommenders have adopted similar acceleration strategies. Existing approaches can be broadly categorized into:

\begin{enumerate}
    \item \textbf{Improved samplers}: DiffCDR \cite{xuan2024diffusion} employs ODE solvers, and DCPR \cite{long2024diffusion} adopts DDIM to speed up sampling.

    \item \textbf{Latent diffusion}: Several works \cite{wu2023diff4rec, xu2024diffusion, cui2024context, yu2023ld4mrec, liu2024toda} perform diffusion in latent representations rather than the original interaction space, reducing dimensionality and computational cost.  

    \item \textbf{Step reduction}: Some methods \cite{jiangzhou2024dgrm, he2024diffusion, li2024multi} simply decrease the number of denoising steps at inference time. 
\end{enumerate}

Despite these efforts, systematic comparisons of inference time against prior state-of-the-art recommender models remain rare. Most works emphasize accuracy improvements without reporting standardized latency benchmarks. For diffusion recommenders to be practically viable, consistent evaluation protocols that jointly measure recommendation quality and inference cost are necessary. In particular, comparisons against lightweight models that excel in real-time serving would clarify the trade-off between performance gains and computational overhead.

Beyond latency, reproducibility and robustness pose additional efficiency-related concerns. Benigni et al. \cite{benigni2025diffusion} report high variance across multiple runs of several diffusion recommender models (DiffRec \cite{wang2023diffusion}, CF-Diff \cite{hou2024collaborative}, GiffCF \cite{zhu2024graph}, DDRM \cite{zhao2024denoising}), with some models underperforming simpler baselines. While stochastic sampling may partially explain this instability, these findings highlight the need for more rigorous experimental protocols and variance reporting.

Recent directions attempt to improve both stability and speed. FlowCF \cite{liu2025flow} adopts flow matching \cite{lipman2023flow} to learn deterministic and straight flow trajectories, enabling faster inference and improved training stability. Other work explores alternative denoising formulations, such as Gaussian mixture modeling \cite{ye2025gaussian}, to accelerate sampling. Additionally, moving beyond unimodal priors toward multi-domain or multi-interest modeling may improve representational efficiency in complex recommendation scenarios.

Overall, advancing diffusion recommenders to deployable systems requires standardized latency benchmarks, robustness evaluation, and principled analysis of the accuracy-efficiency trade-off.

\subsection{Effective Use of Guidance}

Effectively incorporating auxiliary signals into diffusion frameworks remains a key challenge. Original diffusion models \cite{ho2020denoising, song2019generative} were designed for unconditional generation and lack native mechanisms for structured conditioning. Classifier guidance \cite{dhariwal2021diffusion}, later refined as classifier-free guidance \cite{ho2022classifier}, established the dominant paradigm for conditional diffusion. Many diffusion recommender models \cite{wang2024conditional, tomasi2024diffusion, xu2024diffusion, ma2024multimodal, zuo2024diff, jiang2024diffairec} adopt classifier-free guidance to inject auxiliary information into the denoising process. Existing approaches primarily rely on two fusion strategies:

\begin{enumerate}
    \item \textbf{Concatenation}: The noisy representation is concatenated with auxiliary features before being fed into the denoising network.  

    \item \textbf{Cross-attention}: Auxiliary representations interact with the noisy embedding through attention mechanisms.  
\end{enumerate}

While these mechanisms provide effective conditioning, they largely treat guidance as an architectural attachment rather than a principled modeling component. Future work may explore more expressive integration strategies.

More fundamentally, the quality of conditional generation depends on how informative the guidance signal is. Most works directly use raw interaction vectors or encoded features obtained from standard encoders. However, systematically studying how to extract compact, disentangled, or task-aware guidance representations remains largely unexplored. As auxiliary information is often high-dimensional and noisy, identifying which components should influence the generative trajectory is critical.

In sequential recommendation, most models encode historical sequences solely as guidance signals. A few recent approaches instead diffuse the sequence jointly with the target item, creating dual pathways where sequence modeling participates directly in the generative process. This joint diffusion paradigm opens new possibilities for modeling temporal uncertainty and user intent evolution.

Finally, the denoising network itself is often implemented as a simple MLP or Transformer backbone, with limited architectural analysis. Ablation studies rarely investigate how denoiser depth, attention structure, or conditioning configuration affect recommendation quality. Given that the denoiser defines the reverse dynamics, careful architectural design and systematic evaluation are essential for fully leveraging diffusion models in recommender systems.

\subsection{FAccT}

As the majority of papers focus on task performance, responsible recommendation has received comparatively less attention despite its importance. Commonly referred to as FAccT, fairness, accountability, and transparency, these three perspectives provide key lenses for analyzing ethical recommender systems \cite{ekstrand2024facctrec}.

\subsubsection{Fairness}

Fairness seeks to prevent systematic discrimination or disproportionate advantage toward specific user or item groups \cite{wang2023survey}. Research in diffusion-based recommendation is still limited. DifFaiRec \cite{jiang2024diffairec} introduces fairness-aware diffusion objectives, while Malitesta et al. \cite{malitesta2024fair} show that directly adopting diffusion architectures (e.g., DiffRec \cite{wang2023diffusion}) may increase overall system unfairness. Existing studies mainly concentrate on popularity bias \cite{chen2023bias, gupta2024guided, he2024balancing}.

However, diffusion recommenders have not been systematically evaluated under other well-established biases, including selection bias \cite{liu2022rating}, exposure bias \cite{mansoury2024mitigating}, conformity bias \cite{jia2023biasing}, position bias \cite{collins2018position}, inductive bias \cite{chen2021autodebias}, and feedback-loop amplification \cite{mansoury2020feedback}. It remains unclear whether iterative denoising dynamics mitigate or amplify such biases. A unified benchmarking framework is therefore needed to clarify the relationship between generative modeling and fairness outcomes.

\subsubsection{Accountability}

Accountability concerns privacy preservation and robustness to malicious manipulation \cite{bernard2023systematic}. In privacy-preserving recommendation, federated learning aggregates locally trained models without sharing raw data \cite{sun2024survey}. SDRM \cite{lilienthal2024multi} replaces original user data with diffusion-generated synthetic data, representing an initial attempt to integrate diffusion models into privacy-aware pipelines. The efficacy, computational overhead, and privacy-utility trade-offs of such approaches remain largely unexplored.

From a security perspective \cite{himeur2022latest}, diffusion models have already been used as attacker tools. IPDGI \cite{chen2023adversarial} generates adversarial product images to facilitate poisoning attacks \cite{nguyen2024manipulating}, while ToDA \cite{liu2024toda} produces fake user profiles for shilling attacks \cite{si2020shilling}. Other attack vectors, such as fake review generation \cite{duma2024fake}, remain unexplored in diffusion-based settings. On the defense side, GDMPD \cite{yuan2023manipulating} employs diffusion-based purification for attack detection. More comprehensive threat modeling and robust training strategies are needed for generative recommenders.

\subsubsection{Transparency}

Transparency aims to provide explanations and foster user trust \cite{sonboli2021fairness, wang2024trustworthy}. DIEXRS \cite{guo2023explainable} combines diffusion models with a decoder to generate user-item explanations, marking an initial step toward explainable diffusion recommendation. Future work may explore explanation generation directly within diffusion architectures, including diffusion language models \cite{li2022diffusion} and latent diffusion models \cite{lovelace2024latent}.

Beyond explanation generation, diffusion models offer structural opportunities for transparency. Unlike conventional neural recommenders, diffusion models explicitly expose forward and reverse processes. The reverse denoising trajectory gradually transforms noise into high-likelihood item representations, potentially enabling step-wise analysis of how recommendation decisions emerge. Leveraging this iterative structure for process-level interpretability remains an open direction.

\subsection{Large Language Models}

LLMs have shown remarkable success in natural language processing and related tasks, and recent work has applied foundation models such as ChatGPT \cite{brown2020language} and Llama \cite{dubey2024llama} to recommender systems \cite{zhao2024recommender, wu2024survey}. Leveraging LLMs with diffusion-based recommenders provides a promising pathway to enrich item and user representations, guide generation, and improve explainability. LLMs can enhance recommendation in multiple ways:

\begin{enumerate}
    \item \textbf{Item representation:} Models like iDreamRec \cite{hu2024generate} generate detailed textual item descriptions, which are encoded to produce richer embeddings. DeftRec \cite{qu2025generative} uses LLM reasoning on user preferences as guidance during the reverse diffusion process, improving generation without additional model training.

    \item \textbf{User and sequence modeling:} LLMs can produce textual summaries of user interests, sentiment, habits, and temporal trends. Encoding these descriptions can guide diffusion models to capture shifts in user preferences or multi-aspect interests. Some approaches even diffuse historical sequences jointly with the target item, allowing the generative process to leverage sequence dynamics directly.

    \item \textbf{Explainability:} Textual outputs from LLMs can generate human-readable explanations for recommendations. This can be done post hoc or integrated into the diffusion workflow, improving transparency and user trust.
\end{enumerate}

While LLM-based guidance can be precomputed and does not substantially increase computational cost, several research directions remain open. These include studying how to condition diffusion models on multiple aspects of user or item representations, and systematically analyzing how LLM knowledge interacts with diffusion dynamics to maximize recommendation quality. Additionally, exploiting LLMs for richer and interpretable explanations without adding complexity is an important future direction.

\section{Conclusion}
\label{sec:conclusion}
Due to its ability to model complex distributions, diffusion models have recently gained popularity in recommender systems. We present a comprehensive survey on diffusion models in recommender systems to provide a full landscape of such applications and encourage further research efforts. We prepare readers with the relevant background knowledge and present our taxonomy, classifying existing works into three orthogonal axes. By grouping the works based on their recommendation tasks, we enable readers to easily identify the subgroups of their interest. Additionally, we compile all the datasets used in the current literature and describe their properties. Finally, we discuss and describe four open research directions worthy of further efforts to advance the field of recommender systems with diffusion models. As the field is growing, parts of the landscape can be incomplete and in exploration in the meantime. Given a recent surge of interest in this field, we hope to deliver an informative background to readers. Nonetheless, we look forward to reading more papers in this field and further expanding and solidifying the landscape.

\bibliographystyle{ACM-Reference-Format}
\bibliography{survey}

\end{document}